\documentclass[acmsmall,nonacm]{acmart}
\usepackage{graphicx}
\usepackage[normalem]{ulem}
\usepackage{enumitem}
\usepackage{tabularray}
\usepackage{xcolor}
\usepackage{booktabs}
\usepackage{listings}  
\usepackage{caption}
\usepackage{subcaption}
\usepackage{comment}
\usepackage{multirow}
\usepackage{amsmath}
\usepackage{xspace}
\usepackage{float}
\usepackage[linesnumbered,noend]{algorithm2e}
\usepackage[utf8]{inputenc}

\usepackage{balance}
\usepackage{flushend}

\newcommand{\mypara}[1]{\paragraph{{#1}}}

\AtBeginDocument{%
  }

\setcopyright{cc}
\setcctype{by}
\acmDOI{10.1145/3763060}
\acmYear{2025}
\acmJournal{PACMPL}
\acmVolume{9}
\acmNumber{OOPSLA2}
\acmArticle{282}
\acmMonth{10}

\begin{document}

\title{Model-Guided Fuzzing of Distributed Systems}

\author{Ege Berkay Gulcan}
\orcid{0000-0003-1237-0829}
\affiliation{%
  \institution{Delft University of Technology}
  \city{Delft}
  \country{Netherlands}
}
\email{E.B.Gulcan@tudelft.nl}

\author{Burcu Kulahcioglu Ozkan}
\orcid{0000-0002-7038-165X}
\affiliation{%
  \institution{Delft University of Technology}
  \city{Delft}
  \country{Netherlands}
}
\email{b.ozkan@tudelft.nl}

\author{Rupak Majumdar}
\orcid{0000-0003-2136-0542}
\affiliation{%
  \institution{MPI-SWS}
  \city{Kaiserslautern}
  \country{Germany}
}
\email{rupak@mpi-sws.org}

\author{Srinidhi Nagendra}
\orcid{0000-0002-7171-5543}
\affiliation{%
  \institution{IRIF - CNRS - Université Paris Cité}
  \city{Paris}
  \country{France}
}
\affiliation{%
  \institution{Chennai Mathematical Institute}
  \city{Chennai}
  \country{India}
}
\email{nagendra@irif.fr}

\begin{abstract}
We present a coverage-guided testing algorithm for distributed systems implementations.
Our main innovation is the use of an abstract formal model of the system that is used to define coverage.
Such abstract models are frequently developed in the early phases of protocol design and verification 
but are infrequently used at testing time.
We show that guiding random test generation using model coverage can be effective in covering interesting points
in the implementation state space.
We have implemented a fuzzer for distributed system implementations and abstract models written in TLA+.
Our algorithm achieves better coverage over purely random exploration as well as random exploration guided by different
notions of scheduler coverage and mutation.
In particular, we show consistently higher coverage on implementations of distributed consensus protocols such as Two-Phase Commit and the Raft implementations in Etcd-raft and RedisRaft and detect bugs faster.
Moreover, we discovered 12 previously unknown bugs in their implementations, four of which could only be detected by model-guided fuzzing.
\end{abstract}

\ccsdesc[500]{Software and its engineering~Software testing and debugging}
\ccsdesc[500]{Theory of computation~Distributed computing models}

\keywords{Random testing, Fuzzing, Formal models, Distributed systems}

\newcommand{\exec}[1]{\ensuremath{\llbracket #1 \rrbracket}}
\newcommand{\run}[1]{\ensuremath{\llbracket #1 \rrbracket}}
\newcommand{\ls}[1]{\ensuremath{\llbracket #1 \rrbracket}}
\newcommand{\rt}[1]{\ensuremath{\llbracket #1 \rrbracket_{rt}}}
\newcommand{\tuple}[1]{\left<{#1}\right>}
\newcommand{\runs}[1]{\ensuremath{\mathsf{Runs}(#1)}}
\newcommand{\traces}[1]{\ensuremath{\mathsf{Traces}(#1)}}
\newcommand{\floor}[1]{\lfloor #1 \rfloor}

\newcommand{\card}[1]{\ensuremath{\lvert #1 \rvert}}
\newcommand{\set}[1]{{\{ #1 \}}}
\newcommand{\powerset}{\raisebox{.15\baselineskip}{\Large\ensuremath{\wp}}}

\newcommand{\prog}{\ensuremath{\mathcal{P}}\xspace}

\newcommand{\true}{\mathtt{true}}
\newcommand{\false}{\mathtt{false}}

\newcommand{\eventtype}{\mathtt{et}}
\newcommand{\recv}{\mathtt{recv}}
\newcommand{\send}{\mathtt{send}}
\newcommand{\msg}{\mathtt{msg}}
\newcommand{\out}{\mathtt{out}}
\newcommand{\proc}{\mathtt{proc}}
\newcommand{\crash}{\mathtt{crash}}
\newcommand{\restart}{\mathtt{restart}}
\newcommand{\deliver}{\mathtt{deliver}}
\newcommand{\buffer}{\mathtt{buff}}

\newcommand{\Procs}{\mathtt{Procs}}
\newcommand{\Msgs}{\mathtt{Msgs}}
\newcommand{\Buffer}{\mathtt{Buffer}}
\newcommand{\EventTrace}{\mathtt{EventTrace}}

\newcommand{\toolname}{{\textsf{ModelFuzz}}\xspace}

\def\RM#1{\textcolor{red}{\textbf{RM:} #1}}
\newcommand{\srinidhi}[1]{\textit{\textcolor{blue}{Srinidhi: #1}}}
\newcommand{\burcu}[1]{\textit{\textcolor{orange}{Burcu: #1}}}
\newcommand{\ege}[1]{\textit{\textcolor{green}{Ege: #1}}}

\newcommand{\todo}[1]{\textit{\textcolor{gray}{TODO: #1}}}
\newcommand{\update}[1]{\textit{\textcolor{green}{#1}}}

\lstdefinelanguage{scala}{
  morekeywords={abstract,case,catch,class,def,%
    do,else,extends,false,final,finally,%
    for,if,implicit,import,match,mixin,%
    new,null,object,override,package,%
    private,protected,requires,return,output,sealed,%
    super,this,throw,trait,true,try,%
    type,val,var,while,with,yield,%
    Round,ProcessID,Int,send,update,init,old,process,Boolean,Set,Map,variable,interface,receive},
  otherkeywords={=>,<-,<\%,<:,>:,\#,@},
  sensitive=true,
  morecomment=[l]{//},
  morecomment=[n]{/*}{*/},
  morestring=[b]",
  morestring=[b]',
  morestring=[b]"""
}

\lstdefinelanguage{ho}{
  morekeywords={abstract,case,catch,class,def,%
    typedef, struct,%
    int, bool,%
    Snd,
    Upd,
    received,
    do,extends,final,finally, continue,break,%
    for,implicit,import,match,mixin,%
    new,null,object,override,package,%
    private,protected,requires,return,sealed,%
    super,this,throw,trait,try,%
    type,val,var,while,yield,%
    Round,ProcessID,Int,send,update,init,old,process,Boolean,Set,Map,variable,interface,receive, round, UPDATE, SEND, out_internal,exit, HOmachine, in,out},
  otherkeywords={=>,<-,<\%,<:,>:,\#,@},
  sensitive=true,
  morecomment=[l]{//},
  morecomment=[n]{/*}{*/},
  morestring=[b]",
  morestring=[b]',
  morestring=[b]"""
}

\definecolor{dkgreen}{rgb}{0,0.6,0}
\definecolor{gray}{rgb}{0.5,0.5,0.5}
\definecolor{mauve}{rgb}{0.58,0,0.82}

\lstset{frame=tb,
  morekeywords={output},
  extendedchars=true,
  language=scala,
  showstringspaces=false,
  columns=flexible,
  basicstyle={\scriptsize\ttfamily},
  numbers=left,
  numberstyle=\tiny\color{gray},
  keywordstyle=\color{blue},
  commentstyle=\color{dkgreen},
  stringstyle=\color{mauve},
  frame=single,
  breaklines=true,
  breakatwhitespace=true
  tabsize=3,
  emph={ident,pattern,stmnt},
  emphstyle=\itshape,
  escapeinside={<@}{@>}
}

\def\ContinueLineNumber{\lstset{firstnumber=last}}
\def\StartLineAt#1{\lstset{firstnumber=#1}}
\let\numberLineAt\StartLineAt

\maketitle

\section{Introduction}
\label{sec:intro}

Large-scale distributed systems form the core infrastructure for many software applications.
It is well-known that designing such systems is difficult and error-prone due to the interaction
between concurrency and faults, and subtle bugs often show up in production.
Thus, designing testing techniques that cover diverse and interesting program behaviors to find subtle bugs has been an important research challenge.

Coverage-guided fuzzing, which guides test generation toward more coverage, has been effective in exploring diverse executions, mainly in the sequential setting, using structural coverage
criteria as a feedback mechanism~\cite{Heuse_AFL_2022,zeller2019fuzzing}. However, adopting coverage-guided fuzzing for testing distributed system implementations is nontrivial since there is no common notion of \emph{coverage} for distributed system executions. 
Unfortunately, structural code coverage criteria such as line coverage can ignore the orderings of message interactions in a system, thus missing interesting schedules.
On the other hand, more detailed criteria, such as traces of messages, may provide too many coverage goals and thus consider each random
trace a new behavior, giving up the advantages of coverage-guided exploration.

In this paper, we propose to use \emph{state coverage} in an \emph{abstract formal model} of the system as a coverage criterion and present \emph{model-guided fuzzing} of distributed systems.
Abstract formal models are often developed in the design phase of distributed systems to model and formally analyze the underlying
protocols~\cite{DBLP:books/aw/Lamport2002,DBLP:conf/pldi/DesaiGJQRZ13,DBLP:conf/pldi/DeligiannisDKLT15,DBLP:conf/cloud/DeligiannisGLQ21,DBLP:conf/sosp/BornholtJACKMSS21,DBLP:journals/cacm/NewcombeRZMBD15}.
We show that these artifacts are also beneficial in the continuous testing infrastructure of the implementations themselves.
Our experiments show that a formal model can serve as a good ``guide'' for a random testing engine---this is because the formal model
often captures the important scenarios of the protocol, and coverage of states in the model correlates well with coverage of interesting
behaviors in the implementation.

Our motivation for using abstract model guidance to generate test executions shares common insights with semantic fuzzing~\cite{DBLP:conf/issta/PadhyeLSPT19} and grammar-based fuzzing approaches~\cite{DBLP:journals/sigsoft/LePPLVS19,DBLP:conf/sigsoft/GopinathMZ20} used for testing sequential programs.
Semantic fuzzing aims to cover interesting program executions processing program inputs rather than spending exploration budget for exercising uninteresting, syntactic input parsing logic. While a naive fuzzer is likely to generate inputs that cannot pass the input validation and parsing stage, semantic fuzzing generates test inputs that can go deep into the execution. Similarly, a naive event scheduler for distributed systems is likely to produce tests that spend execution budget in exercising uninteresting, network setup stages. For example, it can explore many different orderings of vote messages during the cluster's leader election phase, barely electing a leader after a prolonged execution. Our approach aims to direct testing toward interesting system behaviors, e.g., processing of user requests once a leader is elected. 
Similar to grammar-based fuzzing that uses formal specification of the test input to guide test generation, we use an abstract formal model of distributed systems to guide the generation of semantically interesting temporal event schedules.

At a high level, the abstract models recognize semantically interesting behavior; the use of abstract states is a way to provide coverage
criteria that capture program semantics.
Of course, the use of abstract models is not a panacea: 
the abstraction may not cover certain implementation details where bugs may lurk. 
However, lack of structural coverage after model-guided
exploration can indicate where additional testing effort should focus, 
as well as point out aspects of implementation behavior that are not covered by the model.

We have implemented our algorithm for testing distributed systems implementations using TLA+ models~\cite{DBLP:books/aw/Lamport2002} of protocols.
In a nutshell, our testing algorithm proceeds as follows.
We start by exploring random schedules of messages, but feed the same sequence of messages to the TLA+ model.
We use the TLC model checker \cite{DBLP:conf/charme/YuML99} to generate the set of reachable model states corresponding to the explored schedule.
We mark a schedule as ``interesting'' if it covers a new state of the abstract model.
We perform, as in coverage-guided fuzzing, a \emph{mutation} of an interesting schedule by swapping the receipt order of two randomly chosen messages or changing the processes to crash. 
Applying a mutation to an event schedule gives a new schedule to explore that is similar to the original schedule but likely to exercise new system behavior.

We applied our algorithm to test an implementation of the Two-Phase Commit protocol and two industrial implementations of Raft in Etcd-raft and RedisRaft. Our evaluation shows that model-guided fuzzing leads to higher coverage and can detect bugs faster than pure (unguided) random testing, structural code-coverage guided fuzzing, and trace-based coverage-guided fuzzing. Besides reproducing known bugs, we discovered 12 previously unknown bugs in the implementations of Etcd-raft and RedisRaft. Moreover, four of the new bugs could only be detected by model-guided fuzzing.

Overall, we make the following contributions:
\begin{enumerate}
    \item We propose using abstract models of the systems to guide the testing of system implementations and present \toolname, a model-guided fuzzing approach for distributed systems.
    \item We implement \toolname for testing the two production implementations of the Two-Phase Commit and Raft protocols and evaluate it compared to the existing approaches.
    \item We discovered 12 unknown bugs in total in the implementations of Etcd-raft and RedisRaft.
\end{enumerate}

\section{Overview}
\label{sec:overview}

In this section, we motivate and overview the use of state coverage in an abstract system model to guide test generation on an example distributed system. First, we describe the example system with a concurrency bug in its implementation. Then, we motivate coverage-guided testing to detect such bugs more effectively. Finally, we show that model-guided coverage provides more useful information in guiding test generation than other coverage notions.

\lstset{basicstyle=\ttfamily\small,breaklines=true}

\begin{figure*}[!ht]
 \centering
\begin{minipage}[t]{0.95\textwidth}
 \centering
\begin{minipage}[t]{0.5\textwidth}
\begin{lstlisting}[frame=none,mathescape=true]
class AppMaster: StateMachine {
  StateMachine worker;
  StateMachine terminator;

  def onRegister(Register r) = {
    StateMachine m = r.stateMachine;
    if (m instanceOf Worker)
      worker = m;   
    else if (m instanceOf Terminator)
      terminator = m;  
  }

  def onRequest(Request r) = {
    if (isReady()) {
      worker.send(Execute(r));
      terminator.send(Terminate(worker));    
    }}
  
  def isReady(): Boolean = {
    return worker != null 
      && terminator != null; }
}
 \end{lstlisting}
 \end{minipage}
 \hspace{2mm}
 \begin{minipage}[t]{0.45\textwidth}
\begin{lstlisting}[frame=none,mathescape=true,firstnumber=23]
class Worker: StateMachine {
  Buffer buffer;
  
  def onInit(StateMachine a) = {
    initBuffer(buffer);
    a.send(Register(this)); }
  
  def onExecute(Request r) = {
    // if (buffer != null) 
    runTask(buffer, r); }
  
  def onFlush() = { buffer = null; }
}

class Terminator: StateMachine {
  
  def onInit(StateMachine a) = {
    a.send(Register(this));}
  
  def onTerminate(Worker w) = { 
    w.send(Flush); }
}

 \end{lstlisting}
 \end{minipage}
 \subcaption{An example system with three processes \lstinline{AppMaster}, \lstinline{Worker}, and \lstinline{Terminator}}
 \label{overview:program}
  \end{minipage}
\begin{minipage}[t]{0.99\textwidth}
\begin{minipage}[b]{0.34\textwidth}
 \centering
 \includegraphics[width=0.85\textwidth]{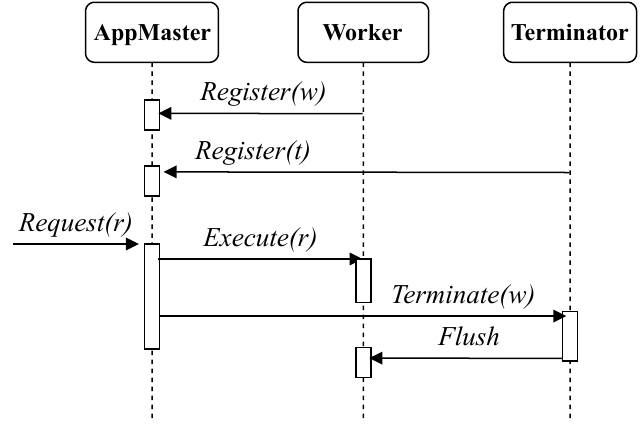}
  \subcaption{A correct execution}
   \label{overview:correct-execution}
  \includegraphics[width=0.85\textwidth]{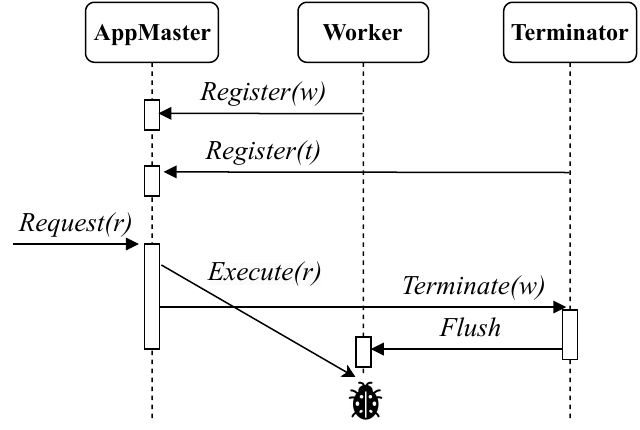}
 \subcaption{A buggy execution}
   \label{overview:buggy-execution}
\end{minipage}%
\hspace{2mm}
\begin{minipage}[b]{0.6\textwidth}
    \centering
    \includegraphics[width=0.95\textwidth]{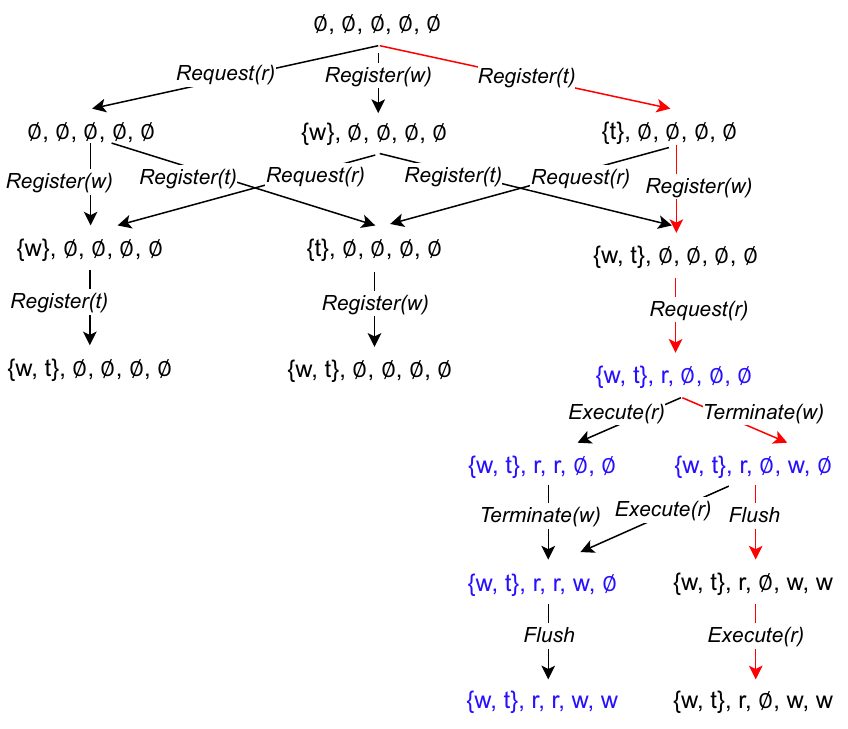}
   \subcaption{Execution space of the system's abstract model}
      \label{overview:execution-space} 
\end{minipage}%
\end{minipage}%
\caption{An example distributed system, its two possible executions, and the execution space of the system's abstract model.} %
\label{fig:motiv-ex}
\end{figure*}

\mypara{An example distributed system.} 
Figure~\ref{overview:program} presents an example system implemented in a Coyote-like actor programming framework~\cite{DBLP:phd/us/Agha85,DBLP:conf/cloud/DeligiannisGLQ21}.
The system consists of three processes \lstinline{AppMaster}, \lstinline{Worker}, and \lstinline{Terminator}.
Each runs in a separate process and communicates with each other by exchanging asynchronous messages. \lstinline{AppMaster} receives client requests, coordinates their execution by a \lstinline{Worker} process, and manages the termination of the worker using a \lstinline{Terminator} process. It accepts the registration messages from \lstinline{Worker} and \lstinline{Terminator} and registers them (lines 5-11). Upon receipt of a client request  \lstinline{Request(r)}, it checks whether the cluster is ready by checking the registrations of the worker and terminator. If it is ready, it sends \lstinline{Execute(r)} to \lstinline{Worker} to handle the request and sends \lstinline{Terminate(w)} to \lstinline{Terminator} (lines 13-17). The \lstinline{Worker} and \lstinline{Terminator} register to the \lstinline{AppMaster} upon their initialization (lines 26, 39). The \lstinline{Worker} handles \lstinline{Execute(r)} by processing it (line 30) and \lstinline{Flush} by cleaning up its buffers (line 34). The \lstinline{Terminator} handles \lstinline{Terminate} by sending \lstinline{Flush} to the worker (line 42). 

\lstset{basicstyle=\ttfamily\footnotesize,breaklines=true}

\lstset{basicstyle=\ttfamily\small,breaklines=true}

The above implementation has a message race bug~\cite{DBLP:conf/asplos/Leesatapornwongsa16}, which occurs in a particular delivery ordering of the asynchronous messages. The worker code mistakenly omits to check if the buffer is null before processing a task (line 31). This causes 
the worker to access a null pointer while processing the client request if \lstinline{Flush} is delivered to it before \lstinline{Execute(r)}. 
Figures~\ref{overview:correct-execution} and~\ref{overview:buggy-execution} illustrate a correct and an incorrect execution of the implementation.

Although the bug seems simple, it is hard to discover such bugs using naive random testing.  Manifesting the bug requires reaching a system state that allows the system to produce \emph{interesting} executions. 
In our example, the processes must register themselves to the \lstinline{AppMaster} before the system serves client requests. 
Thus, pure random testing can be ineffective at exposing distributed system bugs since the generated test cases can get stuck in uninteresting parts of the execution space 
or explore redundant executions.  %

\mypara{Coverage-guided testing} Coverage-guided techniques guide the generation of test executions toward unexplored system behaviors to increase test coverage and search for bugs more efficiently. These methods track the coverage of each test execution and use this information while generating new test cases. 
However, structural test coverage metrics, such as line or branch coverage, are ineffective for assessing the coverage of distributed system executions since they do not capture the concurrency behavior of the distributed systems. 
For example, both correct and buggy executions in Figures~\ref{overview:correct-execution} and~\ref{overview:buggy-execution} hit exactly the same lines and branches of the program in Figure~\ref{overview:program} but with different processing orders of the messages. 

\mypara{Model coverage.} 
Instead, we propose to use an abstract, formal, model of the system 
(e.g., its TLA+ model~\cite{DBLP:books/aw/Lamport2002}) 
to assess test coverage and guide the test generation. %

Figure~\ref{overview:execution-space} shows the execution space of the abstract system model. We consider the $\text{TLA}^+$ model of the system\footnote{\scriptsize https://github.com/burcuku/tlc-controlled-with-benchmarks}, which describes the system's behavior as a transition system specified by a set of states and a set of transitions between them. %
The figure illustrates the possible executions of the system, including the system states and transitions, where each transition corresponds to the delivery of a particular message.
A state of the $\text{TLA}^+$ model of the system is defined by (i) the state of the \lstinline{AppMaster}, which keeps the set of registered processes $\mathit{registered}$ and the set of client requests to process $requests$, (ii) the state of the \lstinline{Worker}, which keeps the set of completed tasks, $\mathit{completed}$, and (iii) the state of the \lstinline{Terminator}, which keeps the set of tasks to terminate, $\mathit{toTerminate}$ and the set of terminated tasks $\mathit{terminated}$. For simplicity, we denote the state of the system as a tuple $\langle \mathit{registered}, \mathit{requests}, \mathit{completed}, \mathit{toTerminate}, \mathit{terminated}\rangle$. The possible actions in the system are the delivery of the messages $\mathit{Request}(r)$, $\mathit{Register}(w)$, $\mathit{Register}(t)$, $\mathit{Execute}(r)$, $\mathit{Terminate}(w)$, and $\mathit{Flush}$, which match the delivery of the messages \lstinline{Request(r)}, \lstinline{Register(w)}, \lstinline{Register(t)}, \lstinline{Execute(r)}, \lstinline{Terminate(w)}, and \lstinline{Flush} in the system's implementation, respectively. 
As given in Figure~\ref{overview:execution-space}, the execution starts in the initial state $\tuple{\emptyset, \emptyset, \emptyset, \emptyset, \emptyset}$ and updates the system state based on the actions taken. 
Our approach assesses the coverage of a set of executions by measuring their \emph{state coverage} in the abstract formal model. 

We could have alternatively considered \emph{trace-based} coverage based on the Mazurkiewicz traces arising out of an execution. Mazurkiewicz traces associate two sequences of messages with the same Mazurkiewicz trace if they only reorder \emph{independent} messages, whose reordering does not affect the system's behavior.  
Our state-based notion is coarser than traces while keeping the essential information about the system's behavior.
For example, although the messages \lstinline{Register(w)} and \lstinline{Register(t)} are dependent, their relative ordering does not affect the system state. 
On the other hand, the ordering of \lstinline{Request(r)} affects the reached system state; the system handles the request only if it is delivered after the two registration messages. The given system has 10 possible message orderings with 8 Mazurkiewicz traces (capturing the commutativity of the \lstinline{Execute} and \lstinline{Terminate} messages). However, the set of all possible system states can be covered by running fewer executions, e.g., only 2 executions for this example. 

\mypara{Model-guided testing.} Now, we show that guiding the generation of test executions using the model state coverage directs the exploration toward interesting behaviors more effectively than with trace-based coverage.
This is because model-state based coverage captures the covered set of program behavior more succinctly than trace-based coverage, which labels each reordering of dependent events as a new coverage class regardless of the system behavior.

Consider an example where the following set of executions has been explored:
\lstset{basicstyle=\ttfamily\footnotesize,breaklines=true}
\begin{itemize}[leftmargin=+.25in]
    \item [\scriptsize{E1}] \lstinline{Request(r)}, \lstinline{Register(w)}, \lstinline{Register(t)}
    \item [\scriptsize{E2}] \lstinline{Register(w)}, \lstinline{Request(r)}, \lstinline{Register(t)}
    \item [\scriptsize{E3}] \lstinline{Request(r)}, \lstinline{Register(t)}, \lstinline{Register(w)}
    \item [\scriptsize{E4}] \lstinline{Register(t)}, \lstinline{Request(r)}, \lstinline{Register(w)}
    \item [\scriptsize{E5}] \lstinline{Register(w)}, \lstinline{Register(t)},  \lstinline{Request(r)}, \lstinline{Execute(r)}, \lstinline{Terminate(w)}, \lstinline{Flush}
    \item [\scriptsize{E6}] \lstinline{Register(t)}, \lstinline{Register(w)},  \lstinline{Request(r)}, \lstinline{Terminate(w)}, \lstinline{Execute(r)}, \lstinline{Flush}
\end{itemize}

\lstset{basicstyle=\ttfamily\small,breaklines=true}

Coverage-guided testing selects the executions that hit previously unseen coverage classes, and it generates new test executions by mutating them. 
Trace-based coverage would label all these executions as \emph{interesting} since each belongs to a different coverage class (i.e., they deliver \lstinline{Request(r)}, \lstinline{Register(w)}, and \lstinline{Register(t)} to the same process in a different order). 
Generating new tests around all of these executions is likely to result in many redundant runs since many of them (E1-E4) already demonstrate the same system behavior.

In contrast, state-based coverage identifies the coverage of new states in the executions of E5 and E6, which hit some new system states (marked blue in Figure~\ref{overview:execution-space}) that are not observed in E1-E4.
Therefore, state-based coverage-guided testing generates new test cases only around these executions. 
A mutation that swaps the order of \lstinline{Execute(r)} and \lstinline{Flush} in the execution (marked by the red arrows in the figure) triggers the concurrency bug in the implementation:

\lstset{basicstyle=\ttfamily\footnotesize,breaklines=true}
\begin{itemize}[leftmargin=+.25in]
\item[\scriptsize{E7}] \lstinline{Register(t)}, \lstinline{Register(w)}, \lstinline{Request(r)},  \lstinline{Terminate(w)}, \lstinline{Flush}, \lstinline{Execute(r)}
\end{itemize}
\lstset{basicstyle=\ttfamily\small,breaklines=true}

In summary, while naive random testing tends to spend the exploration budget exercising the reorderings of the first messages in Figure~\ref{overview:execution-space} and trace-based coverage guidance promotes exploration around all unique traces, our model-based coverage-guided testing directs the execution to the unexplored, interesting executions. %

\lstset{basicstyle=\ttfamily\small,breaklines=true}

\section{Coverage-Guided Fuzzing with Abstract Models}
\label{sec:fuzzer}

We use \emph{coverage-guided fuzzing} for exploring distributed system executions and guiding the test generation using abstract model coverage. 
Our method is \emph{complementary} to the traditional fuzzing methods (e.g., AFL~\cite{url/afl}), which explore the space of \emph{program inputs}, as \toolname explores the space of \emph{event schedules} in distributed systems.
In this section, we define executions and  
explain how we adopt the coverage-guided fuzzing approach for the model-guided exploration of distributed system executions.

\subsection{Executions of Distributed Systems} 
\label{sec:defs}

A distributed system ${S}$ consists of a set of processes that concurrently operate on their local states and communicate by exchanging asynchronous \emph{messages}. The processes are equipped with message buffers that keep the messages sent to that process, and they process the messages in their buffers serially.
Upon handling a message, a process can update its local state and/or send new messages to the processes. \footnote{We consider FIFO message queues that preserve the order of messages between the same sender-receiver pairs, as in distributed system frameworks such as 
P~\cite{DBLP:conf/pldi/DesaiGJQRZ13}, Coyote~\cite{DBLP:conf/cloud/DeligiannisGLQ21}, or Akka~\cite{url/Akka}.}

Let $\Procs$ be the set of processes and $\Msgs$ be the set of all messages exchanged in the system.  
We represent the delivery of a message $\msg \in \Msgs$ to its receiver process by an event $e = \tuple{\recv, \send, \msg}$ for $\recv, \send \in \Procs$, where $\recv(e)$ is the receiver of the message, $\send(e)$ the sender, and $\msg(e)$ is the 
message.   %
Let $\Sigma$ be the set of all message delivery events.
A \emph{state} $s$ of the system is {$s: \langle \Buffer: \Procs \times \Procs \mapsto [\Msgs] \rangle$}  where $\Buffer$ is a map from a sender-receiver pair of processes to the list of messages in their message buffers.  
A transition in the system picks a buffer $\buffer$ and executes the first message $\msg$ in it, i.e., running an event $e=\tuple{\recv, \send, \msg}$. 
Executing the message can lead to the creation of messages $m_i$ sent to $\recv$, i.e., $\send$ may send new messages to some processes upon processing $m$. 
The new state $s'$ is obtained by removing $m$ from $s(\buffer)$ and appending $m_i$ to $s(\Buffer(\send, \recv))$ for each $i$, and we write $s \xrightarrow{e} s'$. 
An \emph{execution} of the distributed system is a sequence
$s_0 \xrightarrow{e_0} s_1 \xrightarrow{e_1} \ldots \xrightarrow{e_n} s_{n+1}$
of states $s_i$ and events $e_i$. %
In addition to delivering messages, we introduce two other types of events $\tuple{\proc, \crash}$, and $\tuple{\proc, \restart}$, which respectively correspond to crash or restart events of a process $\proc$. %
We call the sequence $\tuple{e_0, \ldots, e_n}$ an event \emph{schedule}. %

\subsection{Coverage-guided Fuzzing of Distributed Systems} 

\begin{algorithm} [t]

\DontPrintSemicolon
\SetKwProg{proc}{proc}{}{}
\SetKw{And}{\textbf{and}}

\KwIn{A distributed system $S$}
\KwIn{\textcolor{blue}{An abstract model of the system $M$}}
\KwIn{\textcolor{blue}{A map $\mathtt{map}$ from system events to model actions}}
\KwIn{An initial corpus of \textcolor{blue}{test cases} $T_0$}
\KwOut{The set of explored \textcolor{blue}{test cases} $T$}
\KwOut{The set of \textcolor{blue}{covered abstract states} $totalCoverage$}
\BlankLine

\BlankLine
 \proc{$\mathtt{coverageGuidedFuzzer}(S, M, T_0)$}{
  $T \gets T_0$ ; $totalCoverage \gets \emptyset$\;
  \While{test budget not exceeded}{
     \For{$t \in T$}{
         \textcolor{blue}{$coverage \gets \mathtt{executeAndGetCoverage}(S, M, t)$} \;
         \If{$coverage \cap totalCoverage \neq \emptyset$} {
            $p \gets \mathtt{assignEnergy}(coverage)$ \;
            \For{1 to p}{
                \textcolor{blue}{$T' = \mathtt{mutate}(t)$} ; $T \gets T \cup T'$ \;
            }
         }  
         $totalCoverage \gets totalCoverage \cup coverage$ \;
     }
   } 
 \Return $T$, $totalCoverage$ \;
}
\BlankLine 

 \BlankLine
 \proc{$\mathtt{executeAndGetCoverage}(S, M, t)$}{
    \textcolor{blue}{$events \gets \mathtt{executeConcreteSystem}(S, t)$} \;
    \textcolor{blue}{$actions \gets \mathtt{map}(events)$} \;
    \textcolor{blue}{$states \gets \mathtt{executeAbstractModel}(M, actions)$} \;
    \Return states
 }

\BlankLine

\caption{Coverage-guided fuzzing using abstract models. The statements that differ from traditional fuzzers are highlighted in \textcolor{blue}{blue}.}

\label{alg:fuzzing}
\end{algorithm}

Algorithm~\ref{alg:fuzzing} shows the coverage-guided fuzzing algorithm for generating test inputs~\cite{zeller2019fuzzing,DBLP:conf/ccs/BohmePR16}, which we extended for generating test schedules using coverage-guided fuzzing with abstract models. %

The coverage-guided fuzzing algorithm takes the program under test ($S$) and a set of initial test cases ($T_0$) as inputs. It maintains the set of test cases to explore ($T$) and the total coverage ($totalCoverage$) of the executed test cases. After each test execution (line 5), the algorithm checks if the execution covers new system behavior (line 6). If the test case covers new behavior, it assigns an energy value to the test case proportional to the new coverage it achieves (line 7) to explore more around the executions that cover more new behavior. The algorithm generates new test cases by mutating such executions, adds them to the set of executions to explore (lines 8-9), and updates the total coverage with the newly covered behavior (line 10). 
It terminates when a test budget (e.g., specified by a test duration or number of test cases) returns the explored set of test cases together with the test coverage. 

Our approach adopts coverage-guided fuzzing for the exploration of distributed system executions by (i) generating event schedules rather than program inputs, (ii) defining mutations on the event schedules to obtain a new schedule, and (iii) assessing the test coverage based on the coverage in the abstract system model rather than using traditional code coverage metrics. Along with the implementation of the system under test $S$, our algorithm takes an abstract model $M$ of the system as an additional input, which is used to assess coverage. After running a test schedule on the concrete system implementation $S$, it collects the sequence of executed events (line 13) and maps that sequence of concrete system events into the sequence of abstract model actions (line 14). The event mapper method \texttt{map} is provided by the developer, which simply maps an event in the system implementation (e.g., delivery of a message to a process) to an action in the abstract system model. After mapping the system events to model actions, the algorithm runs these actions on the systems' abstract model (line 15). It collects the set of abstract states covered in the system model and returns that as the coverage information for the test schedule (line 16). 

\paragraph{Event schedules as test cases.}
A test case corresponds to a schedule of distributed system events $\tuple{e_0,\ldots,e_n}$, where $e_i$ %
is either (i) the delivery of a message or (ii) crashing a process or (iii) restarting it. 
We collect the set of initial test schedules ($T_0$) by randomly scheduling the events in the executions, where we select the next event uniformly at random among the set of all enabled events.

\paragraph{The space of mutations.}
Mutations are used to extend the search around interesting schedules that cover new coverage goals. 
They introduce small changes to an event schedule to produce new executions that are similar to the original schedule but likely to exercise new behavior.
A difficulty in mutating event schedules is obtaining a feasible schedule after mutating the event order. An offline shuffle of a sequence is likely to produce an infeasible schedule, %
e.g., the new schedule may not contain all the events that appear in the new execution. %

We represent event schedules by referring to the events using the processes that run them rather than directly referring to the delivered messages. 
We represent a test case as a sequence of abstract events $s=\tuple{\buffer_0:a_0} \ldots\tuple{\buffer_n:a_n}$ with $a_0, \ldots, a_n \in \{\deliver, \crash,$ $\restart\}$. Event $\langle \buffer_i:\deliver \rangle$ delivers the first message in $\buffer_i$ to its recipient process, $\tuple{{\buffer_i:\crash}}$ crashes the recipient process of $\buffer_i$, and $\tuple{{\buffer_i:\restart}}$ restarts it. The test case does not explicitly refer to the messages delivered by the $\deliver$ action (i.e., with certain sender and content) but indirectly refers to them using the message buffers whose messages we deliver. 
Moreover, we parameterize $\deliver$ with the number of messages to deliver, e.g., $\tuple{{\buffer_i:\deliver(n)}}$ corresponds to delivering $n$ messages from $\buffer_i$ to the recipient process.
The indirect representation of the messages helps design mutations that result in feasible event schedules while modifying the order of events.
We use the following mutations to generate new schedules: %
\begin{itemize} [leftmargin=1em]
  \item \texttt{SwapBuffers} randomly selects two schedule indices $i, j$ in $s$, and swaps the scheduling order of $\buffer_i$ and $\buffer_j$,
    \item \texttt{SwapCrashProcesses} randomly selects two schedule indices $i, j$ where the recipient of the $\buffer_i$ is crashed at step $i$ and $\buffer_j$ is crashed at step $j$, and swaps the positions of the crash events (for schedules with a single crashing process, it changes the process to crash), %
    \item \texttt{SwapMaxMessages} randomly selects two schedule indices $i, j$ with message delivery events 
    and swaps $q_i$ and $q_j$, i.e., the number of messages to deliver at these positions.
\end{itemize}

\lstset{basicstyle=\ttfamily\small,breaklines=true}

\subsection{Notions of Coverage}
\label{sec:fuzzer:coverage}
The main challenge in designing a coverage-guided testing method for exploring distributed system executions is to decide whether an execution covers interesting behaviors to guide the exploration around that execution. Unfortunately, existing code coverage metrics, which are designed for measuring the coverage of sequential programs, or Mazurkiewicz traces, which provide a syntactic definition of equivalent behaviors in concurrent programs, are unsuitable for measuring the coverage of distributed system executions. 

\mypara{Code coverage metrics.} The traditional coverage-guided testing algorithms, such as AFL++, use code coverage metrics to check whether an execution covers new behaviors. If the execution of a test input exercises new lines or branches of the program under test, it marks that execution as interesting and generates more test inputs similar to that input.
However, the existing code coverage notions used for testing sequential programs are not suitable for measuring the coverage of behaviors of distributed system executions. The same lines or branches of code can be covered by different orderings of the concurrent events in a system, which may result in a different program behavior (illustrated in Section~\ref{sec:overview}).
\mypara{Mazurkiewicz traces.} A foundational formalization of a concurrent system’s possible executions is Mazurkiewicz traces~\cite{DBLP:conf/ac/Mazurkiewicz86}. 
Traces partition the executions of a concurrent system into a set of equivalence classes (traces) based on the orderings of the concurrent events in an execution. Traces are defined on a set of events and an independence relation, where the independent events are commutative; hence, their order in execution does not affect the program behavior. The executions belonging to the same trace order the dependent events in the same total order, but they can reorder the independent events. Hence, traces partition the event sequences based on their partial order on the dependent events. 

While traces provide a precise formalization for defining the notion of equivalence of executions, defining test coverage classes based on traces is too fine-grained, as
many different traces may exhibit the same behavior from the perspective of the protocol
(illustrated in Section~\ref{sec:overview}).
\mypara{Coverage of abstract model states.} 
\label{sec:cov_def}
Our method uses the coverage of states in the system's abstract formal model 
(e.g., written in TLA+~\cite{DBLP:books/aw/Lamport2002}) to assess the coverage of new system behaviors and guide the test executions. 
The states in the abstract model succinctly represent the system's state, abstracting away uninteresting implementation-level details while keeping the essential information of relevant system states and actions. Furthermore, high coverage over the different model states requires exploring all code paths that handle the different system states. We show in our evaluation that high coverage of the formal system states does not degrade the structural code coverage.

We consider the formal model of a system as a labeled transition system ${M} = \langle Q, I, A, \delta \rangle$ where $Q$ is a set of states, $I$ is the set of initial states, $A$ is the set of actions, and $\delta \subseteq Q \times A \times Q$ is a set of transitions. An action $a \in A$ is enabled at state $q \in Q$ iff $(q, a, q') \in \delta$ for some $q' \in Q$. A run of ${M}$ is a sequence $\rho = q_0 \xrightarrow{a_1} q_1 \ldots \xrightarrow{a_n} q_n$ where $q_0 \in I$ and $(q_i, a_i, q_{i+1}) \in \delta$ holds for all $i$.

    Given an execution of a system $S$ with a set of events $Events$, the formal model of that system $M$, and the mapping $\phi: Events \mapsto A$, \emph{the execution $t$ covers the abstract system states} $q_0, \ldots, q_n$ iff the sequence of actions $\phi(e_1), \ldots, \phi(e_n)$ produce the run $\rho = q_0 \xrightarrow{\phi(e_1)} q_1 \ldots \xrightarrow{\phi(e_n)} q_n$ of ${M}$.

\section{Implementation of \toolname}
\label{sec:impl}

\begin{figure}%
    \centering
    \includegraphics[width=0.7\textwidth]{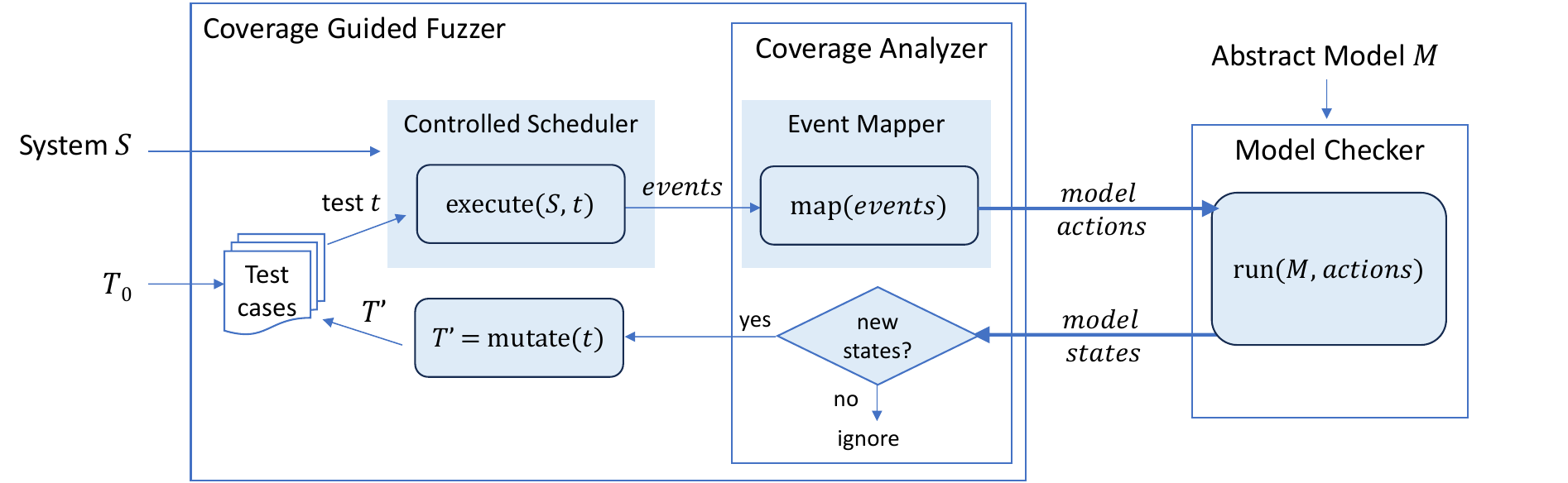}
    \caption{The workflow of model-guided fuzzing}
    \label{fig:workflow}
\end{figure}

\subsection{The Workflow}
Figure~\ref{fig:workflow} illustrates the workflow of model-guided fuzzing. 
Given a distributed system under test $S$ and a set of initial tests $T_0$, the fuzzer runs a test $t$ that represents an event schedule, collects the sequence of executed system events, and maps them to a sequence of actions in the system's formal model. The resulting model actions are run on the system's abstract model $M$. The covered set of abstract model states is used as feedback to guide the test generation.

Different from traditional fuzzers, model-guided fuzzing uses a \emph{controlled scheduler} to control the execution order of events in a distributed system, an \emph{event mapper} to map the system events to abstract actions in the formal model, and a \emph{controlled model checker} to run the given sequence of abstract actions on the model.

\mypara{Controlled scheduler}
The controlled scheduler controls the nondeterminism in the delivery order of messages and executes a particular schedule of events.  
For each event in an event schedule, $t = \tuple{\buffer_0:a_0}\ldots\tuple{\buffer_n:a_n}$ with $a_0, \ldots, a_n \in \{\deliver, \crash,$ $\restart\}$, the controlled scheduler either delivers a message to its recipient process, or crashes a process, or restarts a process.
To enforce the delivery of a message at a certain time in execution, the scheduler intercepts all in-flight messages in the system and collects them in the message buffers for each process. For the process crash and restart actions, we provide scripts for starting and stopping the processes in the system under test. The controlled scheduler runs these scripts to enforce a process crash or restart.

\mypara{Event mapper.}
The event mapper translates the sequence of concrete events executed on the system by the controlled scheduler into a sequence of actions in the abstract model.
Since the implementation level events recorded by the controlled scheduler are specific to the implementation under test, they cannot directly be mapped onto the abstract actions on the formal model. ModelFuzz uses an event mapping interface that the developer implements to map the implementation-level events to the abstract model actions. 
The mapping of the events from the implementation to the model is mostly syntactic, and it does not require an in-depth understanding of the system under test implementation. For example, event mapping for Raft matches the protocol verbs and fields in the collected messages and the TLA+ model actions (e.g., matching the \lstinline{AppendEntries} messages with the receiver process, term, and index numbers).

Our implementation (1) converts the collected implementation-level events to a standard JSON encoding and (2) maps the standardized events in JSON format to abstract actions on the model. 
The second step is embedded in the controlled TLC model checker. The interface to the controlled model checker accepts a sequence of standardized events and outputs the sequence of abstract states observed in the model.

\mypara{Controlled model checker}
The sequence of mapped actions is run on the controlled model checker, which enforces their execution on the model. Unlike a standard model checker, which explores the whole state space of the abstract model, the controlled model checker runs a given sequence of actions and returns the visited set of model states as coverage information. The actions in the specifications used in our evaluation are deterministic, and each execution of the implementation matches a single path in the abstract model. For the models where a single action can lead to multiple model states, model coverage provides an overapproximation, e.g., tracing all paths and collecting all states that could be visited with the executed event schedule. %

For the controlled model checker, we implement a simulation engine that (i) controls the next action to take at each state of the model execution and (ii) records the visited states to provide them as feedback to the fuzzer.
Our implementation uses TLA+~\cite{DBLP:books/aw/Lamport2002} models of the distributed systems for coverage feedback, and we implement the controlled model checker for the TLC explicit state model checker~\cite{DBLP:conf/charme/YuML99} in the TLA+ Toolbox~\cite{DBLP:journals/corr/abs-1912-10633}.

\subsection{Testing Distributed Systems with \toolname} 

\toolname can be used to test any distributed system that has a TLA+ model. To test an implementation of a distributed system using \toolname, two key steps are required: (i) intercepting the messages of the system under test to control their delivery and enforce the delivery ordering specified by the fuzzing algorithm, and (ii) writing an event mapper to map the system's events into the actions defined in the TLA+ model. Testing different system implementations and using different TLA+ models do not require any modifications for the controlled TLC model checker.

The first requirement for controlling delivery is a common requirement across all controlled concurrency testing algorithms. This is essential to enable the execution of specific event schedules generated by the algorithms. Additionally, \toolname requires a TLA+ model of the system as well as an event mapping. TLA+ models are available for popular distributed system protocols, such as Two-Phase Commit, Paxos, and Raft. The event mapping, which maps the implementation events to the abstract TLA+ model actions, is mostly synthetic. It is tailored to the specific system being tested and is provided by the developers.

In the rest of this section, we explain our implementation for testing Raft protocol implementations, providing an overview of the protocol, its abstract model, and the event mapper.

\subsubsection{Testing Implementations of the Raft Protocol}
\label{sec:implementation:raft}

\mypara{Raft protocol.} Raft~\cite{DBLP:conf/usenix/OngaroO14} is a leader-based consensus protocol that proceeds in a sequence of terms. Each term begins with a leader election phase. A \emph{candidate} sends \texttt{RequestVote} messages requesting votes from other processes to be elected as \emph{leader}. A candidate process transitions to the \emph{leader} state if it receives a quorum of accepted votes. Upon successful election, the leader executes client requests, records the executed operations in a log, and replicates the log of operations on the other processes by sending \texttt{AppendEntries} messages to the follower processes. The leader also periodically sends \texttt{Heartbeat} messages to communicate its availability to other processes. If a process does not receive \texttt{Heartbeat} messages from the leader over a time period, it moves to a new term, incrementing its term number, and transitions to the \emph{candidate} state.

\paragraph{Abstract protocol model} We use the TLA+ model of the Raft protocol made available by the protocol's authors~\cite{url/raft-tla}
and extend it by modeling (i) crash and restart of the processes and (ii) snapshot operations. Therefore, besides the variables for the internal states of each process (e.g., the term number, its log of requests), the extended model uses an additional variable to keep the set of active processes in a cluster and introduces crash and restart actions. The crash and restart actions are enabled for the active and crashed processes, respectively, and they update the set of active processes in the cluster. 
Based on empirical evidence that implementation bugs can occur in the snapshot processing logic, we also extend the model to capture snapshot operations so that the model can guide the executions toward executions where processes trigger snapshots and restore them upon recovery.
While the existing TLA+ model does not capture snapshot operations, our model uses a snapshot index variable for each process to record the snapshot index of each process and introduces actions to update the processes' snapshot indices.
Our extended TLA+ model\footnote{https://github.com/burcuku/tlc-controlled-with-benchmarks/blob/main/tla-benchmarks/Raft/model/raft\_enhanced.tla} satisfies the correctness specifications listed in the original model and, therefore, preserves correctness.

\paragraph{Collection of abstract states}
While the abstraction provided by the TLA+ model is useful in guiding the fuzzer, we observe the need for further abstracting the set of observed states to guide the exploration to `interesting' parts of the state space. 
Concretely, the states of the existing TLA+ model are defined by the local states (e.g., operation logs) of each process, along with the current term numbers of each process. Therefore, the model states differentiate between the system states with the same set of local process states if they are reached in different term numbers. Consider the leader election phase of an execution. Although the local states of the processes do not change, unsuccessful leader election rounds result in hitting new system states since the processes increment their term numbers. Such state information guides the fuzzer toward exploring executions with growing term numbers without covering interesting system behavior. 

To guide the fuzzer with more concise state information, we abstract the term numbers in the formal model state. %
Specifically, we merge two consecutive states in an execution, which differ only in terms of the number of non-leaders. 

Note that state abstraction does not require any modification of the formal TLA+ model or the model checker. The abstraction post-processes the collected abstract states from the model checker, and the abstracted states are communicated to the fuzzer. In general, any abstraction can be incorporated by writing an abstraction function from the output state of the model checker to the input state of \toolname.

\paragraph{Event mapper from Raft implementations to the abstract TLA+ actions} 
We implemented event mappers for two different implementations of the Raft protocol (in etcd and Redis) to the abstract TLA+ model.
Our mapper for Raft maps system events collected during the execution into three classes of actions in the TLA+ model: 
\begin{itemize}[leftmargin=+.15in]
    \item The cluster actions \lstinline{AddProcess}, \lstinline{Crash}, \lstinline{Restart}, identified with a process id as an argument: The system events for adding a process, crashing a process, and restarting a process map to the given actions, respectively.
    
    \item The protocol actions of a process, \lstinline{Timeout}, \lstinline{ElectLeader},  \lstinline{UpdateSnapshotIndex}, identified with a process id and protocol arguments (e.g., term and snapshot numbers): The system events to initiate a term change, electing a leader, and updating a snapshot map to the actions, respectively.  

    \item The protocol actions for processing protocol messages, 
    \lstinline{ClientRequest}, \lstinline{HandleRequestVoteRequest}, \lstinline{HandleRequestVoteResponse}, 
    \lstinline{HandleAppendEntriesRequest}, \lstinline{HandleAppendEntriesRequest}, and 
    
    \lstinline{HandleNilAppendEntriesResponse}, 
    identified with protocol arguments (e.g., term and index numbers): The system events of exchanging these messages with corresponding arguments map to the actions.
        
\end{itemize}

\section{Experimental Evaluation}
\label{sec:exp}

We implement \toolname to test two industrial implementations of the Raft protocol~\cite{DBLP:conf/usenix/OngaroO14}: Etcd-raft\footnote{\scriptsize https://github.com/etcd-io/raft} and RedisRaft\footnote{\scriptsize https://github.com/RedisLabs/redisraft} along with an implementation of the Two-Phase Commit Protocol and an implementation of a parametrized version of the example system presented in Section~\ref{sec:overview} in the Coyote framework~\cite{DBLP:conf/cloud/DeligiannisGLQ21}. 
Our implementation for the Raft protocol uses the TLA+ model of the protocol 
made available by the protocol's authors~\cite{url/raft-tla} and extends it\footnote{\scriptsize https://github.com/burcuku/tlc-controlled-with-benchmarks/blob/main/tla-benchmarks/Raft/model/raft\_enhanced.tla} by modeling (i) crash and restart of the processes and (ii) snapshot operations. Our implementation for the Two-Phase Commit protocol is based on the protocol's model available in the TLA+ documentation~\cite{url/tla-doc} and extends it\footnote{\scriptsize{https://github.com/egeberkaygulcan/2PC-Fuzzing/blob/main/tla/TPCL.tla}} with transaction variables and support for multiple requests.

We evaluate the performance of model guidance compared to pure (unguided) random testing, guided testing using structural code coverage and trace coverage information, and, finally, state-of-the-art reinforcement learning guided testing, BonusMaxRL~\cite{DBLP:journals/pacmpl/BorgarelliEMN24}.
For comparison against BonusMaxRL, we leverage the publicly available implementation\footnote{\scriptsize{https://github.com/zeu5/raft-rl-test}}\footnote{\scriptsize{https://github.com/zeu5/dist-rl-testing}} and retain the configurations detailed in the original paper.

We evaluate the performance of \toolname in terms of \emph{test coverage} and \emph{bug finding ability}, answering the following research questions:

\begin{enumerate}[label=\textbf{RQ\arabic*}, leftmargin=3\parindent]
    \item \label{eval:rq1} How does the test coverage of \toolname compare to other strategies?
    \item \label{eval:rq2} Is \toolname more effective at detecting bugs than the other strategies?
\end{enumerate}

We address \ref{eval:rq1} by comparing the abstract state coverage of \toolname to pure random, line coverage-guided, and trace coverage-guided fuzzing strategies. 
We address \ref{eval:rq2} by evaluating the bug-finding effectiveness of different testing strategies in two measures~\cite{DBLP:conf/icse/BohmeSM22}: the unique number of bugs found and the number of test executions to find a bug.

\mypara{Test oracle.} We check the correctness of test executions by checking for assertion violations, exceptions, and crashes. We also check the serializability of the operations in etcd and Redis, running Elle~\cite{DBLP:journals/pvldb/AlvaroK20}, a black-box isolation checker on the collected history of log operations.
on the executed operation history.

\mypara{Statistical evaluation~\cite{DBLP:journals/stvr/ArcuriB14}}.
We analyze the statistical significance of the coverage results of our experiments by performing the Mann-Whitney U-test~\cite{Mann_1947}. We assess \toolname's bug-finding ability compared to the other strategies using Vargha and Delaney's $\hat{A}_{12}$ statistic \cite{vargha2000critique}, with $\hat{A}_{12}=0.6$ as in previous work \cite{DBLP:conf/ccs/MengPRS23}.

\mypara{Test configuration.} 
We run the fuzzers with an initial set of $\card{T_0}=20$ random test cases. For each test case that covers a new state, we create five new test cases by mutating the original test case. We multiply the number of generated test cases proportionally with the number of new states observed in the test execution. 
When the set $T$ of tests to explore becomes empty, we repopulate a random set of tests and repeat the cycle until the test budget is exceeded. %

We run the experiments on an Intel(R) Xeon(R) CPU E5-2667 v2 machine with 32 cores 
and with 252GB of RAM.

\subsection{Microbenchmark in Coyote}
\label{sec:eval:coyote}

\lstset{basicstyle=\ttfamily\small,breaklines=true}

We implemented a parametrized version of the example in Section~\ref{sec:overview} in the Coyote framework~\cite{DBLP:conf/cloud/DeligiannisGLQ21}. 
The implementation parametrizes the system in (i) the number of worker processes and (ii) the number of \texttt{Execute} task messages that need to be processed to handle a request.
For (i), we generate $m$ workers that need to register to the \texttt{AppMaster} before \texttt{AppMaster} can process a client request. 
For (ii), we modify the processing of \texttt{Execute} so that the \texttt{Worker} divides the work into a chain of $n$ number of tasks. 

The system's possible executions involve different interleavings of the \texttt{Terminate} message with the chain of \texttt{Execute} messages sent to the \texttt{Worker}.
We seeded a concurrency bug %
that occurs if \texttt{Terminate} is processed by the \texttt{Worker} just before the last \texttt{Execute} message while processing a client request. 
The bug gets harder to trigger with increasing $m$ and $n$ since it requires all $m$ workers to register to \texttt{AppMaster} before \texttt{Request} and also to deliver the chain of \texttt{Execute} messages except for the last one before \texttt{Terminate}. \footnote{\scriptsize Available at https://github.com/burcuku/coyote-concurrency-testing} %
\begin{figure}[t]
\begin{subfigure}[b]{0.5\linewidth}
    \centering
    \includegraphics[width=1\linewidth]{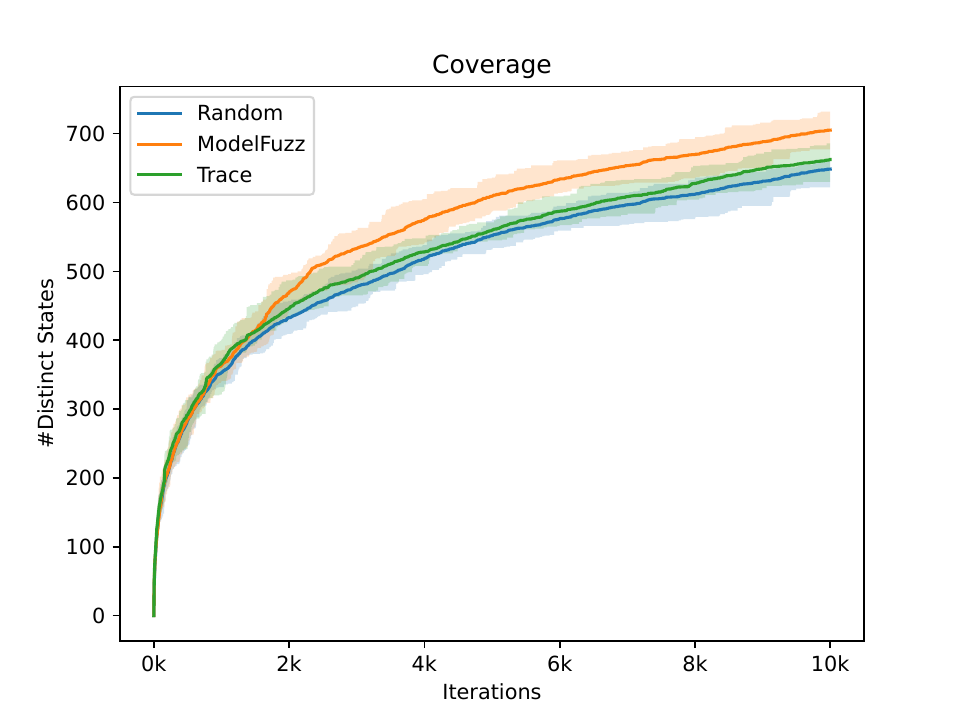}
    \caption{}
    \label{fig:mb-0}
    \end{subfigure}%
    \begin{subfigure}[b]{0.5\linewidth}
        \centering
        \includegraphics[width=1\linewidth]{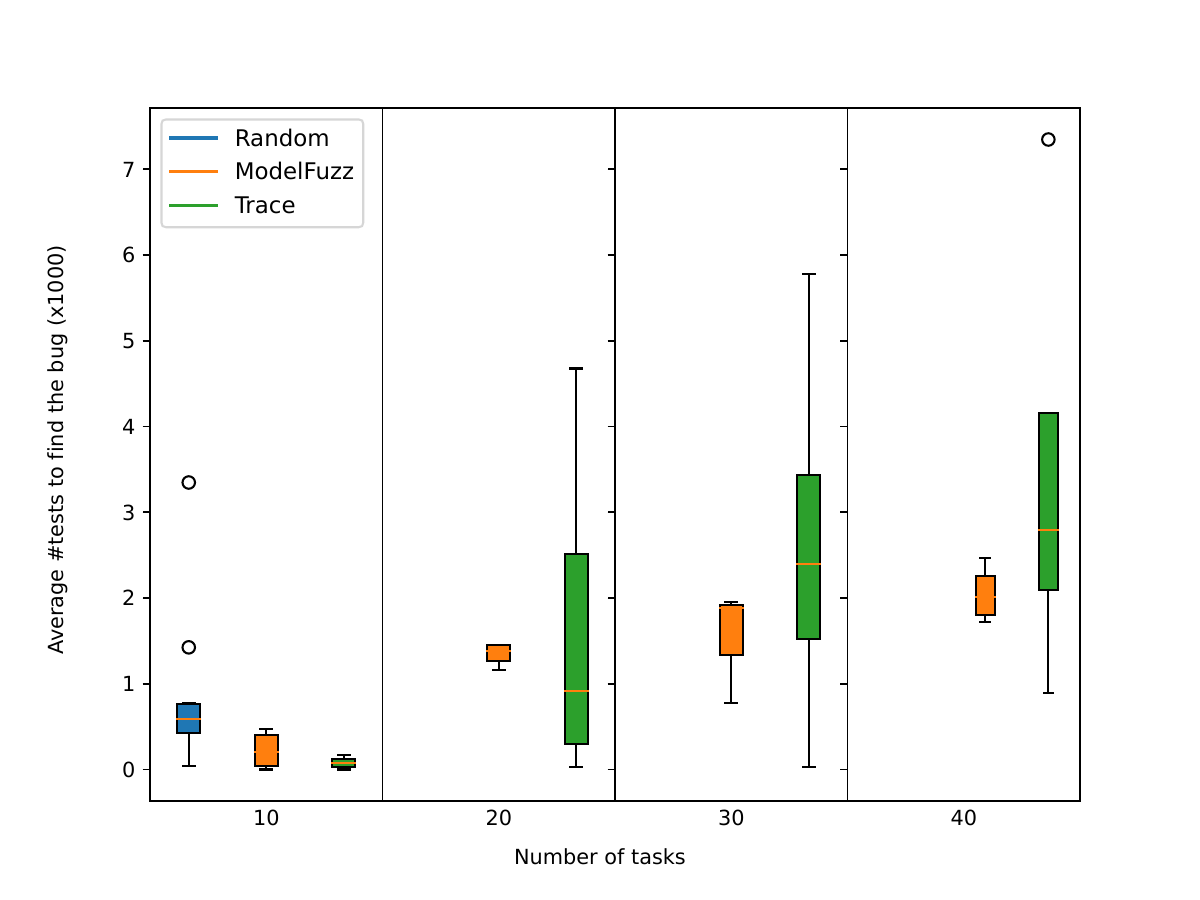}
        \subcaption{} %
        \label{fig:mb-4}
    \end{subfigure}
    \caption{Testing the microbenchmark. (a) Test coverage for $m=6$ workers and $n=40$ tasks (b) \#tests to find the bug for varying \#tasks with $m=6$ workers.}
\end{figure}

\begin{table}[t]
        \setlength{\tabcolsep}{2 pt}
        \begin{tabular}{c | c c c c | c c c c | c c c c }
                      & \multicolumn{4}{c}{$m = 5$} & \multicolumn{4}{c}{$m = 6$}  & \multicolumn{4}{c}{$m = 7$}  \\
          n           & $10$ & $20$ & $30$ & $40$   & $10$ & $20$ & $30$ & $40$   & $10$ & $20$ & $30$ & $40$ \\
        
        \midrule
        
        Random        & \textbf{0.81} & \textbf{1.00} & \textbf{1.00} & \textbf{1.00} & \textbf{0.86} & \textbf{1.00} & \textbf{1.00} & \textbf{1.00} & \textbf{0.71} & \textbf{1.00} & \textbf{1.00} & \textbf{1.00} \\
        
        Trace         & 0.33 & 1.55 & \textbf{0.80} & \textbf{0.83} & 0.32 & 0.43 & \textbf{0.73} & \textbf{0.75} & 0.14 & 0.14 & 0.37 & \textbf{0.60} \\
    \end{tabular}
    \caption{Pairwise $\hat{A}_{12}$ statistic results against \toolname for the microbenchmark with varying $m$ and $n$.} %
    
    \label{tbl:mb-bug-stats}
\end{table}

We ran the microbenchmark with 10K iterations with varying $m=\set{5, 6, 7}$ workers and $n=\set{10, 20, 30, 40}$ tasks over ten runs for each configuration. For this system, we do not compare with line-based coverage as the existing coverage tools do not integrate with the framework we use.

\mypara{Coverage.} Figure \ref{fig:mb-0} shows the coverage of the abstract states of the microbenchmark with $m=6$ workers and $n=40$ tasks, which is representative of different parameter configurations. Since the microbenchmark is a small example with a small state space, %
the difference in the explored number of unique abstract states among different testing approaches is not large. However, the results show \toolname's ability to cover more abstract states compared to random testing and trace coverage guidance. %
Our Mann-Whitney U-tests show that \toolname achieves statistically significantly better coverage results at $\alpha=0.05$, compared to random testing and trace-guided fuzzing with p-values $\set{0.0001, 0.0004}$.

\mypara{Bug finding.} We observe a trend with \toolname where it consistently detects the bug faster than random and trace coverage guided testing approaches with increasing $m$ and $n$. 
Figure~\ref{fig:mb-4} plots the number of test iterations to find the bug for increasing $n$ number of task messages with a fixed $m=6$ processes. The results show that the increasing number of task messages makes the bug more difficult to detect as it is triggered deep in the execution space. This effect is most evidently seen with pure random testing, as it fails to detect the bug after $n=10$. Trace coverage guided testing achieves a more consistent variance among different campaigns. However, we observe a trend in its median value for the first iteration to detect the bug, where it declines as the bug gets harder to detect with increasing $n$. The performance degradation is not as significant for \toolname, as its median value does not change drastically among the experiments. %

We use Vargha and Delaney's $\hat{A}_{12}$ statistic to analyze the significance of our bug-finding results on all parameter configurations. Table~\ref{tbl:mb-bug-stats} lists the pair-wise $\hat{A}_{12}$ statistic values against \toolname for testing the microbenchmark with varying $m$ workers and $n$ tasks. The results show the statistical significance of 17 out of 24 configurations (highlighted in bold), which indicates that \toolname is more effective than the other testing approaches at finding the bug.

\subsection{Two-Phase Commit}
The Two-Phase Commit protocol is a popular distributed algorithm that ensures all nodes in a distributed database either commit or abort a transaction in a coordinated manner. It plays a crucial role in maintaining consistency within distributed databases and transaction management systems. We provide a thread-based implementation, where we modify the standard protocol to handle requests as sets of variables written by a transaction\footnote{\scriptsize https://github.com/egeberkaygulcan/2PC-Fuzzing}. Each node maintains a lock table, which starts in an unlocked state. When a transaction commit request is received, the node checks the lock status of the variables. If none are locked, the vote is approved; otherwise, the request is aborted. 

The implementation omits details such as network communication and storage code, as the benchmark's primary goal is to observe the protocol's behavioral changes under different request and server interleavings. This results in a compact server implementation of 350 lines of Go code, with approximately 150 additional lines for instrumentation. The implementation of \toolname, including the fuzzer loop and communication with the controlled TLC model checker, is approximately 1k lines of Go code.

We tested the Two-Phase Commit protocol using three resource managers (RM), a dedicated transaction manager, two variables (V), and five transaction commit requests (N). These parameters were selected for two key reasons: (i) smaller V values reduce the likelihood of successfully committing concurrent requests, and (ii) the number of distinct abstract model states grows exponentially with increasing N and RM. In experiments with higher N and RM values, we observed similar results, with the only difference being a larger state space requiring a greater time budget for exploration. The chosen parameter set strikes a balance between exploration difficulty and maintaining a manageable state space.

Since the protocol is not fault-tolerant, we excluded crash faults from this benchmark. Each test execution ran for 100 steps per iteration, with a maximum of five messages delivered at each step. We run all tests for one hour and report the average result of 20 test runs.
\begin{figure}[tb]
    \begin{subfigure}[b]{0.44\linewidth}
        \centering
        \includegraphics[width=0.9\linewidth]{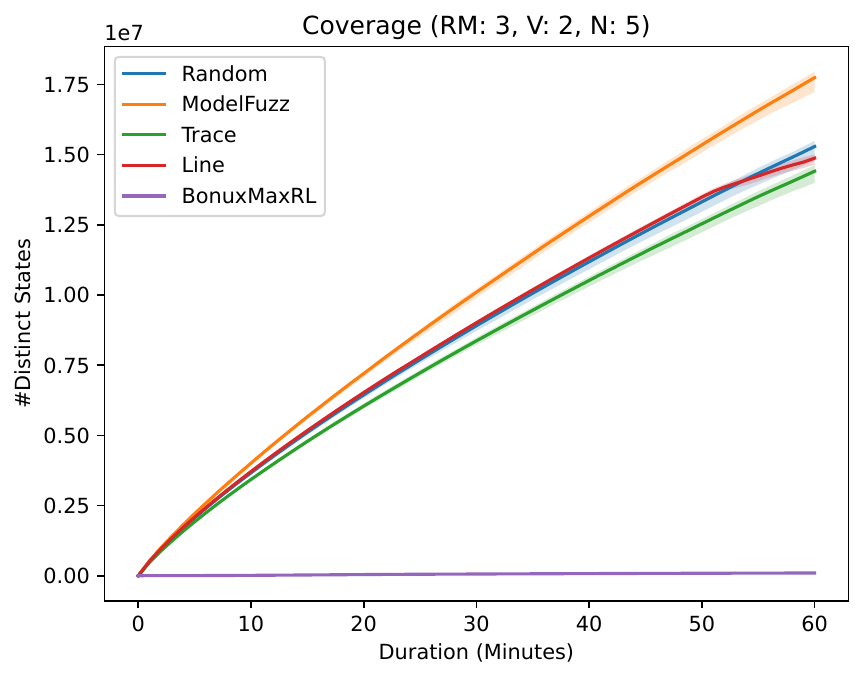}
        \subcaption{} 
        \label{fig:2pc-coverage}
    \end{subfigure}%
    \begin{subfigure}[b]{0.55\linewidth}
    \centering
        \setlength{\tabcolsep}{1.5 pt} 
        \begin{tabular}{l|c}
        Guidance & Avg. \#states  \\
        \midrule
        ModelFuzz & \textbf{17.74M} \\
        Random & 15.29M  \\
        Trace & 14.41M  \\
        Line & 14.87M \\
        RL & 96.98K \\
        \end{tabular}
    \subcaption{}
    \label{tbl:2pc-stats}
    \end{subfigure}
    \label{fig:2pc-results}
    \caption{(a) Test coverage of abstract states of Two-Phase Commit. (b) Average number of abstract states discovered by the strategies. }
\end{figure}
\mypara{Coverage.} Figure~\ref{fig:2pc-coverage} presents the test coverage of the test harnesses, measured by the number of abstract states observed across different strategies. The results show that \toolname surpasses both random and trace coverage-guided exploration strategies. In Figure~\ref{tbl:2pc-stats}, we report the average number of distinct abstract model states discovered by each strategy, with the best result highlighted in bold. Overall, \toolname observes 1.16x more states than random exploration, 1.23x more than trace coverage-guided exploration, 1.20x more than line coverage-guided exploration, and 182.96x more than the reinforcement learning-based testing BonusMaxRL. 

Our analysis of the experiments reveals that BonusMaxRL's poor performance stems from its high learning cost, which exceeds the execution cost of the program. Within the same time budget, fuzzing-based strategies complete three orders of magnitude more iterations than BonusMaxRL. As a result, BonusMaxRL is significantly less efficient at exploring the state space.

Consistent with our other benchmarks, we address ~\ref{eval:rq1} by confirming that model guidance outperforms random exploration, reinforcement learning-guided and other coverage-guidance strategies in the coverage of explored states.

We conduct Mann-Whitney U tests to compare our benchmarks against \toolname and validate the statistical significance of our results. Across 20 experiments, the tests confirm that \toolname discovers significantly more states than the other testing strategies, with p-values of $6.80\mathrm{e}{-8}$ for all comparisons.

\subsection{Etcd-raft}

Etcd-raft\footnote{\scriptsize https://github.com/etcd-io/raft} powers the popular distributed key-value store \textsc{etcd}\footnote{\scriptsize https://etcd.io}. It is a well-tested, production-ready implementation of Raft used by companies such as Cloudflare. %
We instrument its source code~\footnote{\scriptsize https://github.com/zeu5/raft-fuzzing} to intercept the messages exchanged between processes and implement the fuzzer. The highly modular nature of the \textsc{etcd} implementation allows us to instrument and run all the nodes within a single OS process, making it easier and faster to start, stop, and restart all the nodes. The Etcd-raft implementation consists of approximately 7k lines of Go code. We instrument its source code and use an additional 1k lines to implement the fuzzer loop, gain control of the messages exchanged between processes, and implement necessary adapters to communicate with the model checker.

We tested the executions of Etcd-raft with three processes and with five client requests. We ran our tests with a crash quota of 10 and delivered a maximum of 5 messages at each step. For each test run, we execute the test for 12 hours, where each iteration runs for a fixed duration of 1s, and report the average results of 20 test runs.

\begin{figure}[t]
\centering
 \subfloat[\label{fig:etcd-cov-1}]{
   \includegraphics[width=0.5\linewidth]{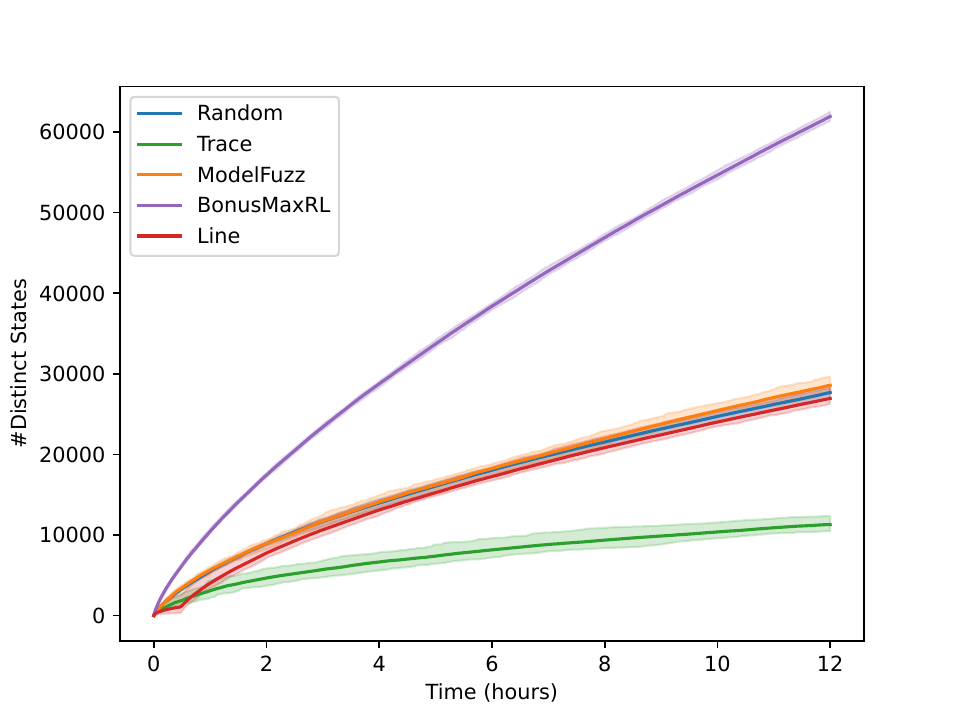}
 } 
 \subfloat[\label{tbl:etcd-bugs}]{
   \small{
   \begin{tabular}{l|c}
        Guidance & Avg. \#iterations \\
        \midrule
        \toolname & 0.243 \\
        Random & 0.143  \\
        Line & 0.322  \\
        Trace & 0.256  \\
        RL & 0.234 \\
        \end{tabular}
        }
 } 
\caption{Testing etcd. (a) Test coverage of abstract states (b) Average number of test iterations to find the synthetic bug. The new bug is only detected by \toolname.}

\end{figure}

\begin{table}[]
    \centering
    \begin{tabular}{c|c|c|c|c}
        Bug & Random & Trace & Line & RL \\
        \midrule
        Seeded & 0.57 & 0.53 & 0.47 & 0.49 \\
    \end{tabular}
    \caption{Pairwise $\hat{A}_{12}$ statistics against \toolname for etcd.}
    \label{tbl:etcd-stats}
\end{table}

\mypara{Coverage.} Figure~\ref{fig:etcd-cov-1} reports the test coverage of the test harnesses 
in the number of abstract states observed with different strategies. The results show that model-guided test generation of \toolname covers similar coverage as random testing (Random), coverage-guided fuzzing using line coverage (Line), and outperforms coverage-guided fuzzing using trace coverage (Trace) in the explored number of unique system states. However, the reinforcement learning approach BonusMaxRL outperforms all other approaches.

Given the nature of our instrumentation with Etcd-raft implementation, \toolname suffers from the performance overhead of communicating with the TLC model checker to measure and retrieve the state trace for the explored execution. However, other approaches are able to achieve better coverage in the absence of this overhead.
Comparing model-guided fuzzing and structural guidance, we find that in both cases, the code coverage saturates at $47.9\%$. Similarly, comparing model-guided fuzzing and trace-based fuzzing, we find that in both cases, we explore 10k unique traces.

\mypara{Bug finding.} Etcd-raft has been the subject of many extensive testing approaches. %
However, we found one new bug in addition to reproducing a seeded bug. The seeded bug modifies the condition for checking if a process has a quorum of votes. Specifically, we change the valid quorum size from $n/2+1$ to $n/3+1$. The new bug we found is more subtle and leads to a process crash when accessing a missing snapshot. 
We reported the bug\footnote{\scriptsize https://github.com/etcd-io/raft/issues/108} to developers.

To answer \ref{eval:rq2}, we analyze the number of detected bugs and the number of test iterations to discover a bug using different strategies. 
While the seeded bug can be found in each of the 20 trials by all of the strategies, the new bug can only be found using \toolname.  
Figure~\ref{tbl:etcd-bugs} reports the average number of iterations to discover the seeded bug using different strategies. The bug can be found by random search faster on average. This can be explained by the characteristics of the bug, which is easily triggered in executions with quorums of only $n/3+1$ processes.

Table \ref{tbl:etcd-stats} compares the distributions of the first occurrence of the seeded bug in each trial against \toolname for the different guidance approaches. 
The Vargha-Delaney ($\hat{A}_{12}$) statistical significance test shows that no approach is significantly better for the seeded bug. 

The results show that \toolname is more effective at finding bugs, as only \toolname finds the new bug, and all approaches show comparable performance for the seeded bug.

\subsection{RedisRaft}

RedisRaft\footnote{\scriptsize https://github.com/RedisLabs/redisraft} powers the popular high-performance Redis distributed key-value store. It compiles into a module that can be loaded onto the main Redis server, which enables different Redis servers to behave as a group and commit client requests in the same order. 
We instrumented RedisRaft to intercept the exchanged messages\footnote{\scriptsize https://github.com/egeberkaygulcan/redisraft-fuzzing} and implement the fuzzer\footnote{\scriptsize https://github.com/zeu5/redisraft-fuzzing}. Overall, the Raft module consists of 30k lines of C code. Our instrumentation adds an additional 1.5k lines of C and Go code, where the Go code implements the fuzzer loop. 
Unlike with Etcd, the instrumentation of RedisRaft requires running all the nodes in different OS processes, along with a network server to capture and deliver messages. Therefore, starting, stopping, and restarting nodes is a time-consuming process.

We tested RedisRaft by running the fuzzer with three processes and five client requests. %
We ran the test executions with a crash quota of 10 and delivered a maximum of five messages at each step. For each test run, we executed the test for 12 hours, each iteration running for a fixed duration of 3s, and reported the average results of 21 test runs. 

\begin{figure}[t]
\centering
 \subfloat[\label{fig:redisraft-cov}]{
   \includegraphics[width=0.5\linewidth]{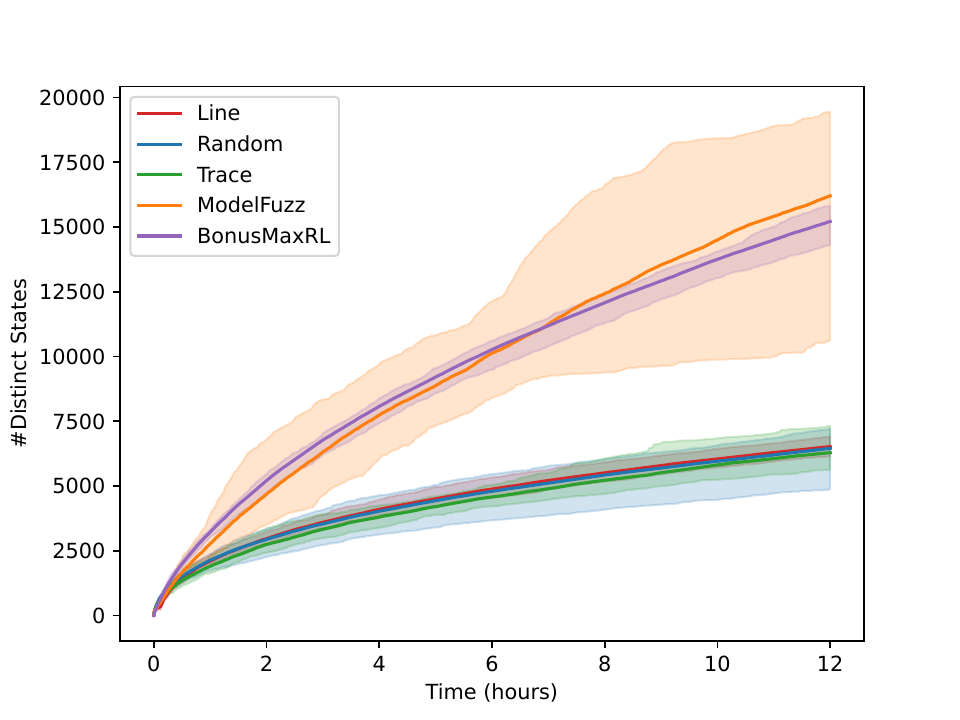}
 } 
 \subfloat[\label{tbl:redis-branch-cov}]{
   \small{
    \begin{tabular}{l|c}
        Method & Branch coverage \\
        \midrule
        \toolname & 149.14 $\pm$ 111.80 \\
        Random & 141.07 $\pm$ 87.36 \\
        Trace & 151.07 $\pm$ 107.94 \\
        Line & 150.64 $\pm$ 97.02 \\
        \\ 
    \end{tabular}
    } 
 }
\caption{Testing RedisRaft. (a) Test coverage of abstract states (b) Average branch coverage of 20 runs.} %
\end{figure}

\mypara{Coverage.} Figure~\ref{fig:redisraft-cov} illustrates the average coverage measures for 12-hour test runs for each testing strategy. We show that \toolname can obtain better coverage over model states compared to random testing, line coverage-guided, and trace coverage-guided fuzzing strategies. The coverage of the reinforcement learning-based approach BonusMaxRL is comparable to \toolname.

On average, \toolname observes 2.58x more states than random exploration, 2.43x more than line coverage, 2.84x more coverage than trace-guided fuzzing, and 1.06x more than BonusMaxRL. We answer ~\ref{eval:rq1} with an observation that model guidance outperforms random exploration and other coverage-guidance strategies in the coverage of explored system states and achieves comparable coverage to the BonusMaxRL strategy.

We perform the Mann-Whitney U test over the final coverage numbers to mitigate the effect of randomness. We obtain p values of $3.12\mathrm{e}{-8}$ vs random, trace, and line coverage guidance strategies. The U tests show that \toolname covers significantly more states than the other traditional guidance strategies, barring the reinforcement learning guided approach. %

We also analyze the branch coverage of the tests and observe that guiding the test executions using model coverage does not degrade the coverage over traditional code coverage metrics. 
Table~\ref{tbl:redis-branch-cov} reports the mean and standard deviation branch coverage of the different guidance methods. 
We measure the branch coverage of the source code in C using the gcov tool. However, we observe a high variance in the data, which we attribute to the coverage measurement tool. As we run multiple copies of the source code during each test iteration, the branch coverage is combined for each copy. However, we found that the gcov tool does not always merge the coverage values of concurrent invocations correctly. Failing to merge coverage information results in the high variance we observe with the branch coverage measures presented in the table.

Unlike for Etcd-raft, \toolname is able to cover more states when compared to other approaches. The difference lies in the nature of the instrumentation. The performance overhead (in time) of communicating with TLC to obtain coverage information is insignificant compared to the time required to restart all the nodes. Therefore, all approaches cover a similar number of iterations, and we are able to achieve better coverage.

\mypara{Bug finding.} Our experiments for testing RedisRaft discovered 13 different bugs, two of which are known bugs reported in RedisRaft's issue tracker, and the remaining 11 are new, previously unknown bugs. The bugs occur in the existence of process crashes and restarts with certain orderings of events, and they manifest as thrown exceptions or assertion violations. We investigated the bugs and reported them in the issue tracker of the RedisRaft open-source repository.\footnote{\scriptsize https://github.com/RedisLabs/redisraft/issues (Issue numbers: \#643-\#649)} Table~\ref{tbl:redis-bugs-desc} briefly describes the new bugs we discovered.

\begin{table*}[t]
    \centering
    \small
    \begin{tabular}{l|p{0.92\linewidth}}
        \toprule
        ID & Bug description \\
        \midrule
        3 & Process crashes when restoring the log from a snapshot stored on disk. \\
        4 & Process crashes when polling peer connections. Specifically, a segmentation failure is raised when reading the connection information of a peer. \\
        5 & After receiving information of a newly added node, the process crashes when setting a flag indicating the node has been successfully added. \\
        6 & Redis server crashes when checking for active client connections. \\
        7 & Process crashes when updating the log after receiving an \texttt{AppendEntries} message. Specifically occurs when it has to delete existing log entries. \\
        8 & Process crashes when sending \texttt{AppendEntries} and reading from a corrupt log.  \\
        9 & Process crashes when updating the state to follower upon receiving a message from the leader. \\
        10 & Process crashes when updating the state to follower upon receiving a message of a higher term. \\
        11 & Process fails to update its current term upon receiving \texttt{AppendEntries} with a higher term. \\
        12 & Process fails to update its current term upon receiving \texttt{RequestVote} with a higher term. \\
        13 & Process fails to update its current term upon receiving \texttt{RequestVoteResponse} with a higher term. \\ 
        \bottomrule
    \end{tabular}
    \caption{The new bugs found in RedisRaft}
    \label{tbl:redis-bugs-desc}
\end{table*}

\begin{figure}[t]
\centering
 \subfloat[\label{tbl:redis-bugs-alt}]{
    \setlength{\tabcolsep}{1.5 pt}
    \begin{tabular}{l|c|c|c|c|c}
        ID & Random & Trace & Line & RL & \toolname \\
        \midrule
        1  & \textbf{0.12}(21)     & 0.18(21)  & 0.15(21)          & -                  & 0.14(21)        \\ 
        2  & 3.72(19)              & 4.26(20)  & 4.06(15)          & \textbf{0.02}(1)   & 4.13(21)        \\ 
        3  & 0.01(21)              & 0.01(21)  & 0.01(21)          & 0.02(21)           & 0.02(21)        \\ 
        4  & 4.84(16)              & 5.00(13)  & \textbf{3.45}(10) & 3.15(3)            & 3.63(12)        \\ 
        5  & 1.78(21)              & 1.35(21)  & \textbf{1.41}(21) & 5.91(8)            & 1.75(21)        \\ 
        6  & -                     & -         & -                 & -                  & \textbf{9.8}(1) \\ 
        7  & -                     & 8.76(2)   & \textbf{5.21}(1)  & -                  & -               \\ 
        8  & 1.94(21)              & 1.82(21)  & \textbf{0.93}(21) & 2.68(19)           & 1.14(21)        \\ 
        9  & 5.76(6)               & 5.05(4)   & \textbf{2.78}(5)  & -                  & 4.71(5)         \\ 
        10 & -                     & -         & -                 & \textbf{0.17}(1)   & 3.89(1)         \\ 
        11 & \textbf{4.67}(1)      & 9.60(1)   & -                 & -                  & 11.41(1)         \\
        12 & -                     & -         & -                 & -                  & 3.93(1)          \\
        13 & -                     & -         & -                 & -                  & 0.51(1)          \\
    \end{tabular}
 }
 \subfloat[\label{tbl:redis-bug-stats}]{
    \setlength{\tabcolsep}{1.5 pt}

    \begin{tabular}{c|c|c|c|c}
        ID & Random & Trace & Line & RL\\
        \midrule
        1 & 0.399 & 0.467 & 0.494 & \textbf{1.000} \\
        2 & 0.491 & 0.540 & 0.521 & 0.000 \\
        3 & 0.434 & 0.506 & 0.471 & 0.585 \\
        4 & 0.529 & \textbf{0.604} & 0.524 & 0.462 \\
        5 & 0.480 & 0.435 & 0.465 & \textbf{0.892} \\
        6 & \textbf{1.000} & \textbf{1.000} & \textbf{1.000} &  \textbf{1.000} \\
        7 & - & - & - & - \\
        8 & \textbf{0.628} & \textbf{0.637} & 0.439 & \textbf{0.667} \\
        9 & 0.583 & 0.5 & \textbf{1.000} & \textbf{1.000} \\
        10 & \textbf{1.000} & \textbf{1.000} & \textbf{1.000} & 0.000 \\
        11 & 0.000 & 0.000 & \textbf{1.000} & \textbf{1.000}  \\
        12 & \textbf{1.000} & \textbf{1.000} & \textbf{1.000} & \textbf{1.000} \\
        13 & \textbf{1.000} & \textbf{1.000} & \textbf{1.000} & \textbf{1.000} \\
    \end{tabular}
 }
\caption{(a) The number of RedisRaft tests for the first occurrences of the
bugs (in hours) using different guidance strategies. (b) Pairwise $\hat{A}_{12}$ statistic results against \toolname.}
\end{figure}

    Among all 13 bugs, \toolname found more bugs than other guidance approaches. Specifically, \toolname found 12 bugs, while random exploration found 8, trace-guided found 9, line-guided found 8, and BonusMaxRL found only 6 of them. 
    For each bug and guidance method, the table in Figure~\ref{tbl:redis-bugs-alt} lists
    the number of runs that find the bug (in parenthesis) and the average first occurrence of the bug (in hours) in the successful runs. For each bug, we highlight the lowest average first occurrence in bold. Among the bugs, one of them is found only using \toolname. Overall, \toolname is more consistent in reproducing the bugs, finding them more frequently than other approaches.

    For each of the bugs, we calculate Vargha and Delaney's $\hat{A}_{12}$ statistics to analyze the pairwise statistical significance of the results. 
    Table \ref{tbl:redis-bug-stats} reports the $\hat{A}_{12}$ statistics against \toolname for each bug, highlighting the results with statistical significance in bold.
    The analysis shows that \toolname detects the bug $\set{6}$ statistically significantly faster than all other guidance approaches, $\set{8,9,10,11}$ statistically significantly faster than some of the approaches, and comparably faster for the remainder.
    While this indicates its ability to detect bugs faster, the $\hat{A}_{12}$ results do not draw clear conclusions on the statistical significance.

    What is significant is the number of bugs and the occurrences of the bugs found by \toolname compared to other approaches. Specifically, while RL based approach BonusMaxRL achieves comparable coverage over the states, the number of bugs found by BonusMaxRL is significantly lesser than \toolname or other traditional guidance measures. The targeted coverage guided fuzzing approach is able to lead the exploration to corner cases leading to better bug finding ability.

\subsection{Comparison to Related Work}

\subsubsection{Comparison to Mallory} A closely related work to our approach is \textsc{Mallory}~\cite{DBLP:conf/ccs/MengPRS23}, a greybox fuzzer for distributed systems that employs reinforcement learning to guide test generation.

While both \textsc{Mallory} and \toolname learn from test executions to guide test generation, they adopt different approaches to learning. Mallory runs a long test execution (with a reported duration of 24 hours in their paper) and learns from the exploration within that test execution. After each action taken during the execution, it updates its reinforcement learning policy and makes decisions about subsequent actions based on that information.
On the other hand, \toolname runs short test executions (each with a duration of a few seconds) and learns from the exploration across executions. After each test, it analyzes the model coverage achieved and generates the next test cases by mutating the previous executions that have led to coverage of new states. 

The two approaches also differ in the coverage information they use to learn from the executions. \textsc{Mallory} uses a timeline abstraction of the system states, which is similar to the trace coverage notion. However, the set of traces is typically too large for effective guidance. \textsc{Mallory} overcomes this problem by allowing the user to manually annotate the system's code to specify the interesting parts and use this information in the learning process. 
On the other hand, \toolname uses the abstract state coverage based on the formal model states (in our experiments, TLA+) of a system.

Our evaluation does not provide an empirical comparison to \textsc{Mallory} across all the benchmark systems we tested, since the application of \textsc{Mallory} requires manual inspection and annotation of the source code under test. Therefore, we focus our comparison on the shared benchmark, RedisRaft. Specifically, we compare the \emph{abstract state coverage} and \emph{bug detection} of the two techniques. 

We ran \textsc{Mallory} using the default parameters in its artifact repository, updating its configuration files for the Docker setup and adjusting for differences in our test platforms. Because \textsc{Mallory}’s standard logs lack the detailed event and state information needed to measure abstract state coverage, we extended its instrumentation to RedisRaft to log this information. A limitation in measuring the abstract state coverage of \textsc{Mallory} tests is due to the length of their test executions. A single test execution of \textsc{Mallory} processes thousands of user requests, as opposed to five client requests in \toolname tests, making it infeasible to run TLC simulations on the event traces. We mitigate this by collecting abstract state information during the test execution, at the same points where \textsc{Mallory} collects its event information.
Since \textsc{Mallory} collects the process logs separately, rather than compiling them into a combined serial log of system events like \toolname does, we merged the process logs of \textsc{Mallory} executions, ordering them by their timestamps. This allowed us to reconstruct the sequence of system states explored during execution. 
We read the logs into the same definition of model state in \textsc{Mallory} as we used for \toolname.

\begin{figure}[t]
\centering
\centering
 \subfloat[\label{fig:modelfuzz-mallory-with-term-abst}]{
   \includegraphics[width=0.45\linewidth]{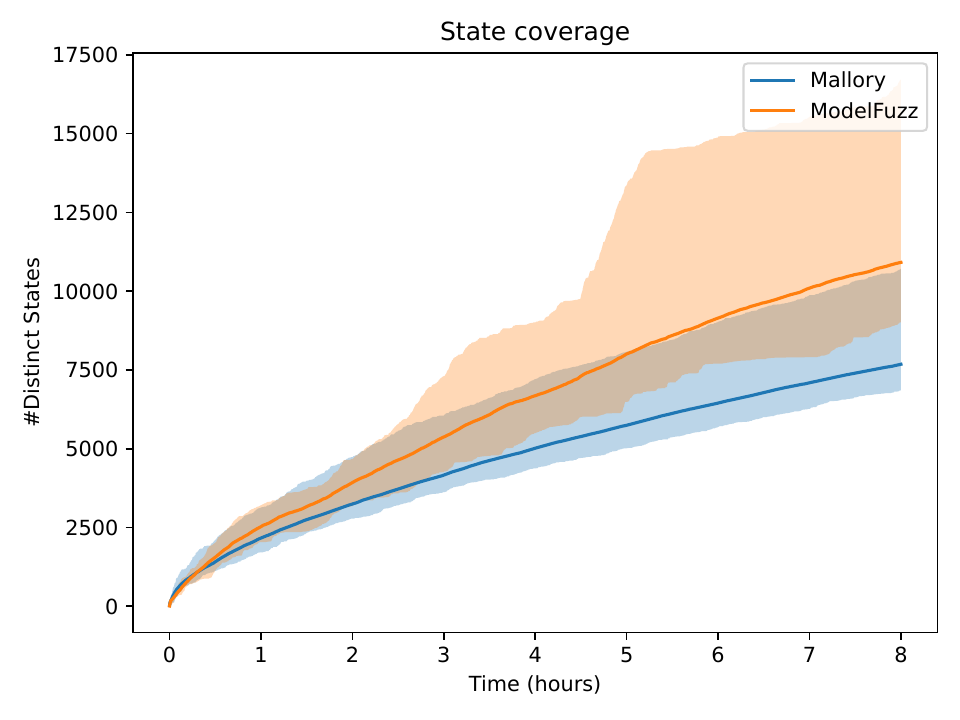}
 } 
 \centering
 \subfloat[\label{fig:modelfuzz-mallory-ignored-active}]{
   \includegraphics[width=0.45\linewidth]{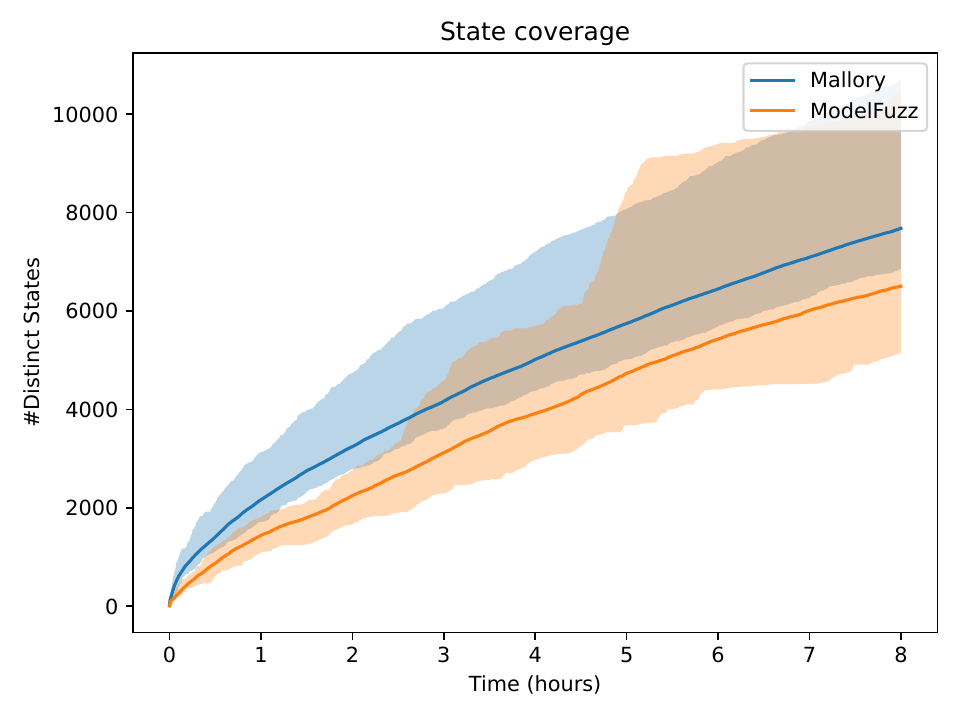}
 } 
\caption{Comparison of abstract state coverage of Mallory and \toolname for testing Redis for 8-hour test runs with the state abstraction of term numbers (a) including the set of active processes (b) excluding the set of active processes in the abstract system state.}
\label{fig:modelfuzz-mallory}
\end{figure}

\mypara{State coverage.} We ran \textsc{Mallory} and \toolname for an 8-hour duration of testing, repeated the tests 20 times, and reported their averages. To ensure a fair comparison in the number of requests processed by \textsc{Mallory} and \toolname, we normalized the logs so that the processes process the same maximum number of user requests simultaneously in the executions of both methods. 

The coverage results showed a linear increase in \textsc{Mallory}'s state coverage over the duration of fuzzing. Upon analyzing the covered states, we observed that the linear growth in coverage is introduced by the increasing term numbers during the protocol execution. The term number is a monotonically increasing counter that tracks the leader election processes in the protocol execution. During the 8-hour test execution of \textsc{Mallory}, the term number increases with the test execution time, while it remains relatively constant, within a few seconds of \toolname tests. To eliminate the effect of the term counter in counting distinct states, we abstracted the term numbers in the state information by considering two states that differ only in the term number as equivalent. 

Figure~\ref{fig:modelfuzz-mallory-with-term-abst} presents the abstract state coverage of \textsc{Mallory} and \toolname with the term number abstraction. 
The results show that \toolname and \textsc{Mallory} perform comparably, with \toolname achieving higher coverage. Figure~\ref{fig:modelfuzz-mallory-ignored-active} presents the state coverage with an additional abstraction of the information of \emph{the set of currently active processes} in the system state. 
As given in Section~\ref{sec:implementation:raft}, we extend Raft's TLA+ model to track the information of crash and restart of the processes and keep that information as part of the abstract system state. \toolname utilizes this information to guide the new tests through the controlled injection of process crashes. 
However, the exact points of process crashes and restarts are not available in the \textsc{Mallory} logs and, therefore, are not read into the abstract states. 
To present a fair comparison based on the same set of collected data, we additionally present the state coverage results without this information in Figure~\ref{fig:modelfuzz-mallory-ignored-active}.
The results show comparable performance for \textsc{Mallory} and \toolname.

\mypara{Bug finding.} 
\toolname tests replicate 2 of the previously known bugs found by both \textsc{Jepsen} and \textsc{Mallory}, and we find 6 new bugs in RedisRaft that were previously not found (described in Table~\ref{tbl:redis-bugs-desc}).
The existence of bugs that could only be detected by \textsc{Mallory} or only by \toolname suggests the complementary bug-finding capabilities of the two methods. \toolname may miss bugs in parts of the implementation (e.g., rolling back a snapshot) that are not captured by the system's abstract model used. %
On the other hand, it can uncover bugs that occur in the parts of the system code that are not user-annotated but captured in the model. %

\vspace{2.5mm}
Overall, \textsc{Mallory} and \toolname show comparable performance in abstract state coverage and bug detection, and they can be used complementarily to each other. The performance of \textsc{Mallory} is influenced by user annotations that identify significant states, while \toolname relies on the expressiveness of the abstract model used.  Depending on the user annotations for \textsc{Mallory} and the abstract model used by \toolname, they can be more effective at covering and detecting bugs in the complementary parts of the execution space. 

\subsubsection{Comparison to BonusMaxRL}
We empirically compare our approach against BonusMaxRL~\cite{DBLP:journals/pacmpl/BorgarelliEMN24}, which utilizes reinforcement learning techniques to explore the state space of a distributed system. Specifically, BonusMaxRL provides a novel reward mechanism that extends prior work and has been proven to be successful in finding bugs in both RedisRaft and Etcd-raft benchmarks. We present empirical comparison of the coverage for all our benchmarks. For the Etcd benchmark, BonusMaxRL outperforms \toolname while for the RedisRaft benchmark, BonusMaxRL achieves comparable coverage to \toolname. However, we observe that BonusMaxRL fails to find all the bugs in both benchmarks.

\subsection{Summary of the Evaluation Results}

In summary, we answer~\ref{eval:rq1}, showing that \toolname outperforms pure random testing and code-coverage and trace-guided fuzzing strategies in the coverage of model states, yet producing coverage comparable to or worse than BonusMaxRL.
We answered~\ref{eval:rq2} by comparing the number of test executions until the first occurrence of a bug. We showed that \toolname discovers bugs faster than the other strategies on average. 
 
 Moreover, the intuition of using model state coverage to guide the fuzzer has shown effective, discovering 1 new bug for Etcd-raft and 11 new bugs for RedisRaft, along with replicating 2 previously known bugs. Note that 2 of the new bugs in RedisRaft discovered by \toolname could not be detected by other approaches
 in the given test budget,
 showing that our approach is more robust in finding bugs earlier than other guidance approaches. 

Finally, our evaluation on several systems under test that implement different distributed protocols demonstrates the applicability of \toolname across multiple systems and the generalizability of the findings.

\section{Limitations}
\label{sec:limitations}

Model-guided fuzzing requires the model to be close to the implementation so that the model provides effective feedback for test generation. The abstraction level of the model heavily affects the performance of model guidance. 
A high-level model abstracting away critical aspects will be ineffective at guiding the tests toward exploring these aspects in the system implementation. 
For example, while the TLA+ model of Raft protocol~\cite{url/raft-tla} %
provided by the protocol's authors models process interactions, it does not capture process crash and restart actions. Therefore, using this model to guide test generation will be ineffective at directing the test executions with process crashes and restarts that potentially lead to previously unseen executions. 
Conversely, a model that is too fine-grained and captures all implementation details is also not helpful. 
For example, a model that implements \texttt{HeartBeat} messages or captures too many details of process states distinguishes equivalent high-level behaviors of the system. 
It can guide test generation toward exploring different executions that produce equivalent system behavior.

\section{Related Work}
\label{sec:related}

\mypara{Coverage-guided fuzzing}
Fuzzing~\cite{DBLP:conf/ccs/BohmePR16,DBLP:conf/kbse/LemieuxS18,zeller2019fuzzing,DBLP:journals/tse/ManesHHCESW21,DBLP:conf/uss/BaBMR22,Heuse_AFL_2022} has been extensively studied for test input generation for sequential programs. 
Rather than generating test inputs independently at random, coverage-guided fuzzing techniques incorporate lightweight program instrumentation to collect some feedback information from the test executions and use that information to guide the generation of new tests. 
Recent techniques target the generation of different types of program inputs~\cite{DBLP:conf/sigsoft/GodefroidHP20,DBLP:journals/corr/abs-2109-11277,DBLP:conf/sigsoft/SteinhofelZ22} and improve fuzzer performance~\cite{DBLP:conf/sigsoft/BohmeMC20,DBLP:conf/issta/ShouTS23}. Extensions of AFL %
~\cite{url/afl} such as \textsc{AFLNet}~\cite{DBLP:conf/icst/PhamBR20}, StateAFL~\cite{DBLP:journals/ese/Natella22} test communication protocols by mutating structured message inputs guided by the states explored. 
Different from test input fuzzing, our work utilizes the fuzzing approach to generate event schedules for testing distributed system executions.

\mypara{Testing distributed systems} Distributed systems
 have been the focus of a wide range of research such as systematic testing and model checking~\cite{DBLP:conf/nsdi/YangCWXLLYLZZ09,DBLP:conf/spin/SimsaBG11,DBLP:conf/sigsoft/DesaiQS15,DBLP:conf/fast/DeligiannisMTCD16, DBLP:conf/osdi/LeesatapornwongsaHJLG14,DBLP:conf/asplos/Leesatapornwongsa16, DBLP:journals/pacmpl/PickDG23, DBLP:journals/pacmpl/EneaGKM24}, random fault injection~\cite{url/jepsen} or scheduling~\cite{url/namazu}, and designing methods for carefully sampling event schedules and faults~\cite{DBLP:conf/cav/ChistikovMN16,DBLP:journals/pacmpl/OzkanMNBW18,DBLP:journals/pacmpl/OzkanMO19,DBLP:conf/asplos/YuanY20,DBLP:journals/pacmpl/DragoiEOMN20,DBLP:journals/pacmpl/WinterBGGO23}. 
While some methods guide fault injection using lineage~\cite{DBLP:conf/sigmod/AlvaroRH15}, runtime state~\cite{DBLP:conf/kbse/ChenDWQ20}, or system's meta-variables~\cite{DBLP:conf/sosp/LuL00TYY19}, most techniques do not exploit information observed in the test executions and generate tests independently from each other.

\mypara{Fuzzing concurrent and distributed systems.}
Fuzzing methods for multithreaded concurrency guide the tests by monitoring races and synchronization events \cite{DBLP:conf/pldi/Sen08,DBLP:conf/icse/WangSG11,DBLP:conf/oopsla/YuNPP12}, execution states caused by thread interleavings~\cite{DBLP:conf/uss/ChenGXSZLWL20}, coverage of concurrent call pairs~\cite{DBLP:conf/ndss/JiangBL022}, and using the reads-from relation between the memory accesses~\cite{conf/asplos/WZDMR24}. These methods are designed for multithreaded programs, and they do not target distributed concurrency.

Recent testing techniques for distributed systems 
learn from the explored executions and adapt reinforcement learning or fuzzing approaches to generate new tests. 
QL~\cite{DBLP:journals/pacmpl/MukherjeeDBL20} uses reinforcement learning to guide the exploration to unexplored parts of the execution space. Recent work~\cite{DBLP:journals/pacmpl/BorgarelliEMN24} improves the reward augmentation using theoretical results in reinforcement learning and programmer-provided information for the interesting parts of the search space. 
Evolutionary search-based testing~\cite{DBLP:conf/icse/MeertenOP23} directs the exploration toward a fitness function.

Existing work for fuzzing distributed systems uses code coverage and message traces as program feedback.
CrashFuzz~\cite{DBLP:conf/icse/GaoDWFWZH23} and FaultFuzz\cite{DBLP:conf/icse/FengP0WD0LL24} fault-injection tools inject process crashes using code coverage information as program feedback. 
\textsc{Mallory}~\cite{DBLP:conf/ccs/MengPRS23} builds on \textsc{Jepsen}~\cite{url/jepsen} %
and guides test generation using programmer annotations that mark interesting parts of the code and an event timeline abstraction (close to traces) as the feedback information. %
Different from using user annotations, code coverage, or trace coverage for guidance, \toolname uses guidance from the abstract system model.

\mypara{Model-based testing}
Model-based testing 
uses formal models (e.g., TLA+ specifications) to exhaustively enumerate system executions and enforce them on the implementation. 
Its applications include testing APIs~\cite{DBLP:conf/hvc/ArthoBHPSTY13}, fragments of HTTP protocol~\cite{DBLP:conf/issta/LiPZ21} and 
specific systems~\cite{DBLP:journals/pvldb/SchvimerDH20}.
Protocol fuzzers DTLS-Fuzzer~\cite{DBLP:conf/icst/Fiterau-Brostean22} and EDHOC-Fuzzer~\cite{DBLP:conf/issta/SagonasT23} use model learning to generate a state machine model of the protocol implementations %
which can be used for model-based testing.

\textsc{Mocket}~\cite{DBLP:conf/eurosys/WangDGWW023} and MET~\cite{DBLP:journals/smr/ZhangHWM24} adopt model-based testing to test distributed systems and CRDTs.
\textsc{Mocket} uses the paths in the model's state space graph as test cases, and it enforces the system under test to run the actions generated on the system's model on the corresponding states and actions in the implementation. For this, it synchronizes the executions of the system and the model at each step. The synchronization of executions requires heavy annotation on the system's source code to mark the program variables and messages associated with the system's model variables. %

Our work conceptually differs from model-based testing, as it performs an unconstrained exploration of the implementation. %
Model-based testing generates test cases using model paths, and hence, it does not cover parts of the implementation that are abstracted away in the model. In contrast, model-guided fuzzing explores the executions of the implementation, including those not captured by the model. 

Recent work~\cite{DBLP:conf/sefm/CirsteaKLM24} uses TLA+ models to validate execution traces of distributed systems. It uses programmer instrumentation on the system implementation to identify the points of state variable updates and transitions and validates the collected execution traces on the TLA+ model to detect the conformance or discrepancies between the system specification and its implementations. 
While this work focuses on verifying execution traces, its future work points to using TLA+ models for guiding test executions towards interesting parts of the search space, as we do with \toolname. 

\mypara{Model-guided fuzzing} \toolname stands out from the state-of-the-art in testing distributed systems as it steers the exploration towards more coverage of system behaviors without the need for the programmer’s comprehension and annotations on the source code (e.g., to identify and mark interesting parts of the source code or using heavy instrumentation to synchronize its execution with a formal model). Instead, it utilizes the system's abstract model, which already captures the essential information about system behavior, to collect information about the explored system behaviors and generate new tests.

\section{Conclusion}
\label{sec:conc}

In this paper, we present \toolname, a new approach for fuzzing distributed systems implementations. 
Our novel approach uses coverage over abstract model states as feedback for the fuzzer to generate test executions and guide the test generation toward interesting parts of the state space. %
We use \toolname to test the implementations of different distributed protocols, including two production distributed system implementations. 
We show that \toolname can achieve high model coverage, allowing us to discover new bugs and replicate known bugs more quickly than other guidance methods.

\section*{Data Availability Statement}
The implementation of \toolname and the systems under test are available in the accompanying artifact ~\cite{artifact/modelfuzz}. 

\begin{acks}
We would like to thank the anonymous reviewers for their constructive comments and suggestions. We would also like to thank Ruijie Meng for her help with running the Mallory tool.
This work was partially supported by the Amazon Research Award.
\end{acks}

\bibliographystyle{ACM-Reference-Format}
\bibliography{references}

\end{document}